\RequirePackage{ifpdf}
\ifpdf
\else
\PassOptionsToPackage{dvipdfmx}{graphicx}
\PassOptionsToPackage{dvipdfmx}{hyperref}
\fi
\documentclass[%
 reprint,
 superscriptaddress,
 amsmath,amssymb,
 aps,prx
]{revtex4-2}

\usepackage{graphicx}
\usepackage{dcolumn}
\usepackage{bm}
\usepackage{hyperref}

\usepackage{amsmath,bm}
\usepackage{subfigure}
\usepackage{slashed}
\usepackage{mathrsfs}
\usepackage{color}

\newcommand{\mtx}[1]{\begin{pmatrix} #1 \end{pmatrix}}

\begin{document}

\title{
Spin systems as quantum simulators of quantum field theories in curved spacetimes
}

\author{Shunichiro Kinoshita}
\email{kinoshita@neptune.kanazawa-it.ac.jp}
\altaffiliation{ present affiliation: Kanazawa Institute of Technology, 7-1 Ohgigaoka, Nonoichi, Ishikawa 921-8501, Japan }
\author{Keiju Murata}
\email{murata.keiju@nihon-u.ac.jp}
\author{Daisuke Yamamoto}
\email{yamamoto.daisuke21@nihon-u.ac.jp}
\affiliation{Department of Physics, College of Humanities and Sciences, Nihon University, Sakura-josui,
Tokyo 156-8550, Japan}
\author{Ryosuke Yoshii}
\email{ryoshii@rs.socu.ac.jp}
\affiliation{Center for Liberal Arts and Sciences, Sanyo-Onoda City University, Yamaguchi 756-0884,
Japan}
\affiliation{International Institute for Sustainability with Knotted Chiral Meta Matter (WPI-SKCM2), Hiroshima University, Higashi-Hiroshima, Hiroshima 739-8526, Japan}

\begin{abstract}
We demonstrate that a quantum field theory (QFT) in general two-dimensional curved spacetimes can be realized by a system of quantum spins or qubits. We consider a spin-1/2 model on a one-dimensional ring with spatially and temporally varying exchange couplings and magnetic fields. This model reduces to a QFT of Majorana fermions in the continuum limit. From this correspondence, we establish a dictionary for translating between the spacetime-dependent parameters of the spin model and the general metric on which the QFT is defined. 
After addressing the general case, we consider the Friedmann-Lema\^{\i}tre-Robertson-Walker (FLRW) metric as a simple example. According to the dictionary, the QFT of Majorana fermions on the FLRW metric corresponds to the Ising model with a time-dependent transverse magnetic field. We demonstrate that the production of Majorana particles in the expanding universe can be simulated with the transverse-field Ising model by increasing the strength of the magnetic field. Furthermore, 
we examine the Unruh effect through the spin system by using our prescription and show the direct relation between the entanglement (or modular) Hamiltonian in the spin system and the Rindler Hamiltonian. 
This approach provides an experimentally viable system for probing various phenomena in QFT within curved spacetime, while also opening the door to uncovering nontrivial phenomena in spin systems inspired by curved spacetime physics. It offers fresh perspectives on both QFT in curved spacetimes and quantum many-body spin systems, revealing profound connections between these fields.
\end{abstract}


\maketitle

\section{Introduction}
Quantum field theory (QFT) in curved spacetimes provides a framework for understanding the behavior of quantum fields in the presence of strong gravitational fields~\cite{BD, Muk}. It is crucial for understanding quantum phenomena in cosmology and black hole physics. For example, QFTs in curved spacetimes give a prediction on quantum fluctuations in an inflationary universe, which produce the primordial inhomogeneities and result in the large-scale structure of the current universe~\cite{Stewart:1993bc}. 
Also, they predict that spontaneous particle production occurs near a black hole and then the black hole can emit a thermal spectrum of particles, namely the Hawking radiation~\cite{Hawking:1975vcx}.
This black hole evaporation due to the Hawking radiation raises 
the information loss paradox, which is one of the biggest unresolved problems in theoretical physics. 
QFT in curved spacetime would offer deep insights into the unification of quantum mechanics with general relativity.

Curved spacetimes play an essential role in gravitational physics, but their experimental realization is difficult because gravity is too weak. 
If the equivalent of curved spacetime could be realized even in non-gravitational systems such as condensed matter systems, it is useful for the analysis of gravitational phenomena, especially its kinematic properties. 
In fact, such an approach has been discussed variously from classical and quantum perspectives as analogue gravity~\cite{Unruh:1980cg,Visser:1997ux,Jacobson:1998ms} (see~\cite{Barcelo:2005fc,Jacquet:2020bar} and the references therein).
For example, propagations of acoustic waves in a perfect fluid can be described by the massless Klein-Gordon equations in an effective curved spacetime.
The effective geometry, so called acoustic metric, depends on the background flows (density, velocity of flow, and local speed of sound) in addition to a physical spacetime metric. 
In the case of transonic flow of fluid, acoustic waves in the supersonic region cannot propagate in a direction opposite to the flow.
The supersonic region corresponds to the black hole region, and the critical point between subsonic and supersonic flows is the horizon.
Furthermore, if we consider quantum excitations in such a situation, we can simulate various phenomena related to the quantum theory of curved spacetime.
Thus, the effective geometry, in various non-gravitational systems: ordinary or superfluids, dielectric mediums, Bose-Einstein condensates, can mimic interesting phenomena of gravitational physics, such as black holes, ergoregion, expanding universe, Hawking radiation, and so on.

In this paper, we highlight how a quantum field theory (QFT) in curved spacetimes can be realized using a system of quantum spins or qubits. Recent advancements in quantum computing have significantly enhanced our ability to simulate complex quantum systems. While current quantum processors are making remarkable progress, they are still evolving, with ongoing improvements in error correction and fault tolerance, moving towards fully scalable and general-purpose computing~\cite{RevModPhys.95.045005}. In this transitional phase, quantum processors are exceptionally well suited for simulating physical systems, offering valuable and practical applications of their capabilities~\cite{Houck2012-hc,Daley2022-ok}. For instance, superconducting quantum processors have demonstrated their ability to perform some calculations beyond the reach of classical approximation methods~\cite{Kim2023-nz}, and they have also shown the capability to simulate many-body phenomena that are challenging for conventional condensed matter systems, such as time crystals~\cite{Frey2022-mp,Mi2022-pl} and measurement-induced phase transitions~\cite{Google_Quantum_AI_and_Collaborators2023-bc}. Additionally, other platforms, such as optical lattice atomic gases~\cite{Gross2017-rb}, Rydberg atom arrays~\cite{Weimer2010-fu,Henriet2020-yk}, and trapped ions~\cite{Barreiro2011-cp,Blatt2012-fa,Monroe2021-tn}, are advancing in their ability to simulate complex spin systems. 
Indeed, the Ising model with transverse and longitudinal magnetic fields, for instance, is already realized by using the optical lattice~\cite{Simon:2011}. 
By leveraging these technologies, we can use various quantum systems to model intricate phenomena in curved spacetimes, bridging the gap between theoretical models and experimental insights. 
This approach offers a powerful means to explore QFTs under various spacetime geometries and showcases the exciting potential of quantum technology.
Note that 
single particle dynamics in a black hole spacetime 
has recently been performed on a superconducting chip~\cite{Yang2020-yc,Benhemou2023-qh,Shi2023-bw}. 

Towards the tabletop quantum simulations of curved spacetimes, we reveal a concrete correspondence between the metric of the most general form of two-dimensional spacetimes and the parameters of a system of quantum spins $S=1/2$ on a one-dimensional ring. By showing that the spin-1/2 model reduces to a QFT of Majorana fermions in the continuum limit via the Jordan-Wigner transformation, we provide a ``dictionary'' for translating between spin and gravity languages. 
Here, supposing the capabilities of current quantum platforms, we consider spatially and temporally varying exchange couplings and magnetic fields in the spin model to establish the correspondence with respect to both space and time coordinates. This approach enables the simulation of time-dependent gravitational phenomena, such as cosmic expansion and gravitational collapse. 

As a typical example, we discuss the quantum production of Majorana particles in the expanding universe modeled with the Friedmann-Lema\^{\i}tre-Robertson-Walker (FLRW) metric. According to our dictionary, this can be simulated using the transverse-field Ising model with time-increasing field strength. We demonstrate that the QFT prediction is well reproduced in systems with up to hundreds of spins or qubits, which aligns with the current capabilities of quantum processors~\cite{Kim2023-nz}. Additionally, we provide a clear interpretation of the Unruh effect from the perspective of the spin model, based on the exact expression of the entanglement Hamiltonian of the transverse-field Ising chain. Our general dictionary offers a solid foundation for quantum simulations of gravitational phenomena across a wide range of quantum platforms, and the demonstrations of several specific but important examples serve as crucial benchmarks for future studies.

This paper is organized as follows.
Section~\ref{sec:Marjorana_2D} is devoted to a quantum field theory of a Majorana fermion in general two-dimensional curved spacetimes.
We write down the Hamiltonian density for the Majorana fermion in terms of a complex variable satisfying the canonical anti-commutation relations.
In section~\ref{sec:spin_model}, we introduce quantum spin chain models with inhomogeneous couplings and transverse magnetic fields.
Applying the Jordan-Wigner transformation, we have a Hamiltonian quadratic in a fermion operator.
We compare the spin chain Hamiltonian in the continuum limit with the field theory of the Majorana fermion and identify free parameters in both models.
In section~\ref{sec:particle_production}, as one of the applications of our results, we consider particle productions in an expanding universe.
We exhibit the number of particles produced by quantum particle production during cosmic expansion for both models: discrete spin model and continuous field theory. In section~\ref{sec:unruh}, we investigate the Unruh effect using our approach through a spin system and demonstrate a direct connection between the entanglement Hamiltonian of the spin system and the Rindler Hamiltonian. The final section is dedicated to the summary and discussion.

\section{Majorana fermions in two-dimensional curved spacetimes}
\label{sec:Marjorana_2D}

We will introduce a QFT of Majorana fermions in two-dimensional curved spacetimes. 
In the next section, we consider continuum limits of general spin chain systems and find that 
some of them are equivalent to the QFT of Majorana fermions.

In the following, we will adopt the $(-+)$-signature for the two-dimensional metric $g_{\mu\nu}$. The gamma matrices in the Majorana representation can be written as
\begin{equation}
\begin{split}
    \gamma^0&=i\sigma^y=
    \begin{pmatrix} 
    0 & 1 \\
    -1 & 0
    \end{pmatrix}\ ,\quad 
    \gamma^1=\sigma^z=
    \begin{pmatrix} 
    1 & 0 \\
    0 & -1
    \end{pmatrix}\ ,
    \\ 
    \gamma^3 &\equiv \gamma^0\gamma^1=-\sigma^x=
    \begin{pmatrix} 
    0 & -1 \\
    -1 & 0
    \end{pmatrix}\ .
\end{split}
\end{equation}
They satisfy $\{\gamma^i,\gamma^j\}=2\eta^{ij}=2 \textrm{diag}(-1,1)$ and $\{\gamma^i,\gamma^3\}=0$ ($i=0,1$). We also introduce zweibeins $e^i_\mu$ satisfying~\footnote{
We use Latin indices $i, j, k, \ldots$ and Greek indices $\mu, \nu, \rho, \ldots$ as indices for the local Lorentz coordinates and the general coordinates, respectively.
}
\begin{equation}
 g^{\mu\nu} e^i_\mu e^j_\nu = \eta^{ij}\ .
\end{equation}

The action for the Majorana fermion with mass $m$ in two-dimensional spacetimes is~\cite{Polyakov:1981re,deLacroix:2023uem,Erbin}
\begin{equation}
    S=-i\int d^2x \sqrt{-g}\, \bar{\psi} (\slashed{\nabla}-m) \psi , \label{eq:Majorana_action}
\end{equation}
where $\bar{\psi}=\psi^\dagger\gamma^0$ and $\psi=(\psi_1,\psi_2)^T$ are a two-component spinor field of real Grassmann variables, satisfying $\psi_a^\dagger = \psi_a$ $(a=1,2)$~\footnote{This condition implies $\psi = \psi_c \equiv C\bar\psi^T$, where $C$ denotes the charge conjugation matrix. In the present representation, $C$, satisfying $C^{-1}\gamma^i C = -(\gamma^i)^T$, is given by $C=\gamma^0$.} and $g$ stands for the determinant of the matrix $g_{\mu\nu}$. The Feynman slash notation is defined as $\slashed{A}=\gamma^i A_i = \gamma^i e_i^\mu  A_\mu$.
We have also introduced the covariant derivative for a spinor field in the zweibein formalism as 
\begin{equation}
   \nabla_\mu \psi = \left(\partial_\mu +\frac{1}{4}\omega_{\mu ij} \gamma^{[i}\gamma^{j]}\right)\psi \ ,
\end{equation}
where $\omega_{\mu ij}$ is the spin connection. 
In the case of two dimensions, however, the term of the spin connection vanishes in the action. This is immediately shown from the Majorana flip relation: $\bar{\chi}\gamma^{i_1}\cdots \gamma^{i_n}\psi = (-1)^n \bar{\psi}\gamma^{i_n}\cdots \gamma^{i_1}\chi$, where $\chi$ and $\psi$ are Majorana fermions.
As the result, the Lagrangian density for Majorana fermions is given by 
\begin{equation}
\begin{split}
    \mathcal{L}=&-i \sqrt{-g}\, \psi^T \gamma^0 (\slashed{\partial}-m) \psi\\
    =&i \sqrt{-g}\big[ e_0^\mu(\psi_1  \partial_\mu \psi_1 +\psi_2  \partial_\mu \psi_2) \\
    &+  e_1^\mu(\psi_1 \partial_\mu \psi_2 +\psi_2 \partial_\mu \psi_1)+m(\psi_1 \psi_2-\psi_2 \psi_1)\big]\ .
\end{split}
\end{equation}
Introducing complex variables, 
\begin{equation}
    \chi=\psi_2-i\psi_1\ ,\quad \chi^\dagger=\psi_2+i\psi_1\ ,
\end{equation}
we obtain 
\begin{equation}
\begin{split}
    \mathcal{L}=&\sqrt{-g}\Big[ \frac{i}{2}e_0^\mu(\chi^\dagger  \partial_\mu \chi +\chi  \partial_\mu \chi^\dagger) \\ &- \frac{1}{2}e_1^\mu(\chi \partial_\mu \chi -\chi^\dagger \partial_\mu \chi^\dagger) +m\chi^\dagger \chi \Big]\ .
\label{Lag0}
\end{split}
\end{equation}

In two dimensions, the most general metric is given by 
\begin{equation}
    ds^2=-\alpha(t,x)^2dt^2+\gamma(t,x)^2(dx-\beta(t,x)dt)^2 .
    \label{generalmetric}
\end{equation}
We introduce zweibeins and their dual vectors as
\begin{equation}
\begin{split}
 &e^0_\mu dx^\mu =\alpha dt\ ,\quad 
    e^1_\mu dx^\mu =\gamma (dx -\beta dt)\ ,\\
&e_0^\mu \partial_\mu =\frac{1}{\alpha}\partial_t + \frac{\beta}{\alpha}\partial_x\ ,\quad 
    e_1^\mu \partial_\mu =\frac{1}{\gamma}\partial_x \ .
\end{split}
\label{zwei0}
\end{equation}
Note that $e^0_\mu dx^\mu$ is taken to be normal to $t$-constant slices.
Then, the Lagrangian density~(\ref{Lag0}) is written as
\begin{multline}
    \mathcal{L}=\frac{i}{2}\gamma(\chi^\dagger  \partial_t \chi +\chi \partial_t \chi^\dagger) 
    +\frac{i}{2}\beta\gamma (\chi^\dagger  \partial_x \chi +\chi \partial_x \chi^\dagger) 
    \\
    - \frac{\alpha}{2}(\chi \partial_x \chi -\chi^\dagger \partial_x \chi^\dagger) +m \alpha \gamma\chi^\dagger \chi \ .
\end{multline}
By introducing a canonically normalized field as 
\begin{equation}
\chi = \gamma^{-1/2} e^{i\zeta /2} \Psi,
\label{Psidef}
\end{equation}
the Lagrangian density becomes 
\begin{equation}
\begin{split}
    \mathcal{L}=&\frac{i}{2}(\Psi^\dagger  \partial_t \Psi +\Psi \partial_t \Psi^\dagger)
    +\frac{i\beta}{2} (\Psi^\dagger  \partial_x \Psi +\Psi \partial_x \Psi^\dagger) \\
       &+ \left[m\alpha - \frac{1}{2}(\partial_t \zeta + \beta \partial_x\zeta)\right] \Psi^\dagger \Psi \\
       &
    - \frac{\alpha}{2 \gamma}\cos\zeta (\Psi \partial_x \Psi -\Psi^\dagger \partial_x \Psi^\dagger)\\
       &
    - i \frac{\alpha}{2 \gamma}\sin\zeta (\Psi \partial_x \Psi + \Psi^\dagger \partial_x \Psi^\dagger) 
    \ .
    \label{PsiLag}
\end{split}
\end{equation}
Note that $\zeta(t,x)$ in the phase of the complex variable $\Psi$ is an arbitrary real function, which corresponds to a rotation of the real spinor field $\psi$ in the action (\ref{eq:Majorana_action}) as 
\begin{equation}
    \psi = 
    \begin{pmatrix}
        \cos(\zeta/2) & - \sin(\zeta/2) \\ \sin(\zeta/2) & \cos(\zeta/2)
    \end{pmatrix}
    \psi' \ .
\end{equation}
This is just a field redefinition, but we will leave it for the convenience of later use.

Finally, we obtain the Hamiltonian density for a Majorana fermion in general two-dimensional metric as 
\begin{equation}
\begin{split}
\mathcal{H}=&-\frac{\alpha}{2 \gamma}\cos\zeta (\Psi^\dagger \partial_x \Psi^\dagger-\Psi \partial_x \Psi) \\
       &
+ i\frac{\alpha}{2 \gamma}\sin\zeta (\Psi^\dagger \partial_x \Psi^\dagger + \Psi \partial_x \Psi)
\\
&  -\frac{i\beta}{2} (\Psi^\dagger  \partial_x \Psi +\Psi \partial_x \Psi^\dagger)\\
       &
-\left[m \alpha - \frac{1}{2}(\partial_t \zeta + \beta\partial_x\zeta) \right]\Psi^\dagger \Psi \ .
\end{split}
\label{MajH}
\end{equation}
The sign of the mass $m$ is not physically relevant since it can be flipped just by the change of variables 
$\Psi\to -i\Psi^\dagger$, $\Psi^\dagger \to i\Psi$ and $\zeta\to -\zeta$. (In terms of the original variables $\psi_a$, this transformation corresponds to a swap of $\psi_1$ and $\psi_2$.)
The field $\Psi$ satisfies the canonical anti-commutation relations 
\begin{equation}
\begin{split}
    &\{\Psi(t,x),\Psi^\dagger(t,y)\}=\delta(x-y)\ ,\\ 
    &\{\Psi(t,x),\Psi(t,y)\}=\{\Psi^\dagger(t,x),\Psi^\dagger(t,y)\}=0\ .
    \label{AntiCanRel}
\end{split}
\end{equation}
See appendix~\ref{CanQ} for the detail of the canonical quantization of Majorana fermions. 
We also give a comment on the natural choice of local Lorentz frame associated with zweibeins there.

\section{Relation between spin models and quantum field theories in curved spacetimes}
\label{sec:spin_model}
\subsection{Spin models and their continuum limits}

In the previous section, we have introduced a QFT for Majorana fermions in a general curved spacetime. 
We will demonstrate that some spin models reduce to the QFT of Majorana fermions in continuum limits.
We consider an $L$ spin model on a one-dimensional ring 
with general couplings $J^{ab}_j(t)$ in the XY plane and out-of-plane magnetic fields $h_j(t)$, both of which are dependent on time and site position:   
\begin{equation}
 H=- \sum_{j=1}^{L} \bigg[\sum_{a,b=+,-} J^{ab}_j(t) \sigma_j^a \sigma_{j+1}^b+ h_j (t) \sigma^z_j\bigg]\ ,
\label{Hgeneral}
\end{equation}
where $\sigma_j^\pm \equiv (\sigma^x_j \pm i\sigma^y_j)/2$ and $\sigma_j^{x,y,z}$ are the Pauli matrices acting on $j$-th site. Periodicity is imposed as 
$\sigma^\pm_{L+1}=\sigma^\pm_1$. 
The hermiticity of the Hamiltonian requires $J_j^{++}=(J_j^{--})^\ast$, $J_j^{+-}=(J_j^{-+})^\ast$ and $h_j\in \bm{R}$.
This spin model is a quantum spin chain with general XY-exchange, Z-component Dzyaloshinskii-Moriya, and Z-component $\Gamma$ interactions (XY-DM-$\Gamma$ model) in the presence of Z-magnetic fields.
The open spin system can be regarded as a special case of the closed system once we set the exchange coupling to zero at a site. (For example, $J^{ab}_{L/2}=0$.) Thus, we can consider both closed and open spin systems simultaneously. 

The spin operators $\sigma^\pm_j$ and $\sigma^z_j$ are written in terms of fermionic operators $c_j$ and $c_j^\dagger$ via the Jordan-Wigner transformation as
\begin{equation}
\begin{split}
    \sigma^+_j=&\prod_{l=1}^{j-1}(1-2c_l^\dagger c_l) c_j\ ,\quad 
    \sigma^-_j=\prod_{l=1}^{j-1}(1-2c_l^\dagger c_l) c_j^\dagger\ ,\\ \sigma_j^z=&1-2c_j^\dagger c_j\ ,
    \label{JW}
\end{split}
\end{equation}
where $c_j$ and $c_j^\dagger$ satisfy the canonical anti-commutation relations $\{c_j,c_l^\dagger\}=\delta_{jl}$ and $\{c_j,c_l\}=\{c_j^\dagger,c_l^\dagger\}=0$. The terms in the Hamiltonian~(\ref{Hgeneral}) are rewritten as
\begin{equation}
\begin{split}
    &\sigma^+_j \sigma^+_{j+1}=c_{j+1}c_j\ ,\quad 
    \sigma^+_j \sigma^-_{j+1}=c_{j+1}^\dagger c_j\ ,\\
    &\sigma^-_j \sigma^+_{j+1}=c_j^\dagger c_{j+1}\ ,\quad 
    \sigma^-_j \sigma^-_{j+1}=c_j^\dagger c_{j+1}^\dagger\ 
\end{split}
\end{equation}
for $j\leq L-1$, and
\begin{equation}
\begin{split}
    &\sigma^+_L \sigma^+_{1}=-e^{i\pi N} c_{1}c_L\ ,\quad 
    \sigma^+_L \sigma^-_{1}=-e^{i\pi N} c_{1}^\dagger c_L\ ,\\
    &\sigma^-_L \sigma^+_{1}=-e^{i\pi N} c_L^\dagger c_{1}\ ,\quad 
    \sigma^-_L \sigma^-_{1}=-e^{i\pi N} c_L^\dagger c_{1}^\dagger\ 
\end{split}
\end{equation}
for $j=L$. Here $N=\sum_{l=1}^L c_l^\dagger c_l$ is the total number operator. The operator 
\begin{equation}
    e^{i\pi N}=\prod_{l=1}^L (1-2c_l^\dagger c_l)=\prod_{l=1}^L \sigma^z_l\ 
    \label{oddity}
\end{equation}
measures the oddity of $N$. One can confirm that $e^{i\pi N}$ is the conserved quantity: $[H,e^{i\pi N}]=0$. 
We now decompose the total Hilbert space $\mathcal{H}$ into $\mathcal{H}_\textrm{even/odd}$ composed of states with the even/odd fermion number as $\mathcal{H}=\mathcal{H}_\textrm{even}\bigoplus \mathcal{H}_\textrm{odd}$.
If we restrict the Hilbert space to $\mathcal{H}_\textrm{even}$ or $\mathcal{H}_\textrm{odd}$, the Hamiltonian is simply written as 
\begin{multline}
 H=- \sum_{j=1}^{L}[ J^{++}_j(t)\, c_{j+1}c_j
 +J^{+-}_j(t)\, c_{j+1}^\dagger c_j\\
 +J^{-+}_j(t)\, c_j^\dagger c_{j+1}
 +J^{--}_j(t)\, c_j^\dagger c_{j+1}^\dagger
 + h_j (t) (1-2c_j^\dagger c_j)]\ ,
\label{Hgeneralc}
\end{multline}
with imposing the periodicity $c_{L+1}=\mp c_1$, where the upper and lower signs correspond to $\mathcal{H}_\textrm{even}$ and $\mathcal{H}_\textrm{odd}$, respectively.
Hereafter, we will drop the $c$-number terms like as $-\sum_j h_j(t)$ in the Hamiltonian.

Let us consider the continuum limit of the Hamiltonian~(\ref{Hgeneralc}). 
We introduce a spatial coordinate of the $j$-th spin as 
\begin{equation}
    x_j = \varepsilon \left(j-\frac{L}{2}\right) ,
\end{equation}
where $\varepsilon$ is the lattice spacing. 
The total physical length of the spin system is given by $\ell = L\varepsilon$. 
Note that the periodicity implies $x_{L+1} = x_1 + \ell \sim x_1$ for closed systems.
If we take the limit of $\varepsilon\to 0$ or $L\to \infty$ keeping $\ell$ finite, we obtain a theory in the continuum limit.
We define the fermionic field by 
\begin{equation}
    \Psi(x_j)=\frac{c_j}{\sqrt{\varepsilon}} ,
    \label{Psi_cnt}
\end{equation}
which satisfies $\{\Psi(x),\Psi^\dagger(x')\}=\delta(x-x')$ in the limit of $\varepsilon\to 0$.
Then, for example, the operator $c_{j}^\dagger c_{j+1}$ in Eq.~(\ref{Hgeneralc}) is written as
\begin{equation}
\begin{split}
    c_{j}^\dagger c_{j+1} &= \varepsilon \Psi^\dagger(x_j) \Psi(x_j+\varepsilon)\\ &= \varepsilon \Psi^\dagger(x_j) \Psi(x_j) + \varepsilon^2 \Psi^\dagger(x_j) \partial_x \Psi(x_j) + \mathcal{O}(\varepsilon^3) \ .
\end{split}
\end{equation}
At the second equality, we assume 
\begin{equation}
    \varepsilon \partial_x \sim \varepsilon k \ll 1\ ,
    \label{lowk}
\end{equation}
where $k$ is a typical wave number~\footnote{
The long-wavelength limit of the spin system does not imply the low-energy limit in general. 
For example, in the XY-model, the gap closes twice in the momentum space and the Dirac fermion appears instead of the Majorana fermion in the low-energy limit.
In this paper, we will focus on the long-wavelength limit and consider the Majorana field.
}. 
Since the physical system size is given by $\ell$, the typical wave number should satisfy $k\gtrsim 1/\ell$.
Similarly, we can express all terms in Eq.~(\ref{Hgeneralc}) using $\Psi(x_j)$. 
We require that $J^{ab}_j(t)=\mathcal{O}(\varepsilon^{-1})$, $h_j(t)=\mathcal{O}(\varepsilon^{-1})$ and $J^{+-}_j(t)+J^{-+}_j(t)-2h_j(t)=\mathcal{O}(\varepsilon^{0})$. 
In terms of their limiting behaviors, we define the real functions $v(t,x)$, $w(t,x)$, $p(t,x)$, $q(t,x)$ and $r(t,x)$  as follows: 
\begin{equation}
\begin{split}
    &2J^{--}_j(t)\varepsilon \to v(t,x_j)+iw(t,x_j)\ ,\\
    &2J^{++}_j(t)\varepsilon \to v(t,x_j)-iw(t,x_j)\ ,\\
    &2J^{-+}_j(t)\varepsilon \to p(t,x_j) + i q(t,x_j) \ ,\\ 
    &2J^{+-}_j(t)\varepsilon \to p(t,x_j) - i q(t,x_j)\ ,\\
    &J^{+-}_j(t)+J^{-+}_j(t)-2h_j(t)\to r(t,x_j)\ ,
\end{split}
\label{Jlim}
\end{equation}
in the continuum limit $\varepsilon \to 0$.

The Hamiltonian in the continuum limit is now given by
\begin{widetext}
\begin{equation}
\begin{split}
    H&=-\int^{\ell/2}_{-\ell/2} dx \bigg[\frac{v(t,x)}{2} (\Psi^\dagger  \partial_x \Psi^\dagger -  \Psi \partial_x \Psi ) 
    +\frac{iw(t,x)}{2} (\Psi^\dagger  \partial_x \Psi^\dagger + \Psi \partial_x \Psi ) \\
 &\hspace{2cm}+\frac{i}{2} q(t,x)
 (\Psi^\dagger \partial_x \Psi +  \Psi \partial_x \Psi^\dagger) 
 +\frac{p(t,x)}{2}
 (\Psi^\dagger \partial_x \Psi -\Psi \partial_x \Psi^\dagger) + r(t,x)
\Psi^\dagger  \Psi\bigg]\\
&=-\int^{\ell/2}_{-\ell/2} dx \bigg\{\frac{v(t,x)}{2} (\Psi^\dagger  \partial_x \Psi^\dagger -  \Psi \partial_x \Psi ) 
    +\frac{iw(t,x)}{2} (\Psi^\dagger  \partial_x \Psi^\dagger + \Psi \partial_x \Psi ) \\
 &\hspace{2cm}+\frac{i}{2} q(t,x)
 (\Psi^\dagger \partial_x \Psi +  \Psi \partial_x \Psi^\dagger)
+ [r(t,x)
-\frac{1}{2} \partial_x p(t,x)
] \Psi^\dagger  \Psi\bigg\} .
\end{split}
\label{Hcontlim}
\end{equation}
\end{widetext}
At the second equality, we have performed integration by parts on the terms involving $p(t,x)$. 
The spatial coordinate $x$ is compactified as $x\sim x + \ell$ for closed systems. If we consider the infinite space, we need to take the limit of $\ell\to \infty$, additionally.

Comparing this Hamiltonian with Eq.~(\ref{MajH}), we can see that the Majorana fermion in the curved spacetime is realized in the continuum limit of the spin model by identifying the functions as
\begin{equation}
\begin{split}
    v(t,x)=&\frac{\alpha(t,x)}{\gamma(t,x)}\cos\zeta(t,x) \ ,\\ w(t,x)=& - \frac{\alpha(t,x)}{\gamma(t,x)}\sin\zeta(t,x) \ ,\quad
    q(t,x)=\beta(t,x)\ ,\\
    r(t,x)-&\frac{1}{2} \partial_x p(t,x) \\=& m \alpha(t,x)  - \frac{1}{2}(\partial_t\zeta(t,x) + \beta(t,x)\partial_x\zeta(t,x))\ .
\end{split}
\label{vw_rel}
\end{equation}
Note that these equations do not uniquely determine all functions in the spin model even if all functions in the field theory are given.
In particular, we can take 
the combination of the two functions, $r(t,x) + \frac{1}{2} \partial_x p(t,x)$,
as an arbitrary function. 
This degree of freedom means that the same field theory can be realized from a number of spin models with different parameters.
From Eqs.~(\ref{Jlim}) and (\ref{vw_rel}), we can explicitly write the parameters of the spin model (\ref{Hgeneral}) in terms of metric components as
\begin{align}
    &J^{++}_j(t)=(J^{--}_j(t))^\ast =\frac{\alpha(t,x_j)}{2\varepsilon\gamma(t,x_j)}e^{i\zeta(t,x_j)}\ ,\\
    &J^{-+}_j(t) = (J^{+-}_j(t))^\ast=\frac{p(t,x_j)+i\beta(t,x_j)}{2\varepsilon}\ ,\\
    &h_j(t)=\frac{p(t,x_j)}{2\varepsilon}-\frac{1}{4}\partial_x p(t,x_j) -\frac{m\alpha(t,x_j)}{2}\notag\\
    &\hspace{0.8cm}+ \frac{1}{4}(\partial_t\zeta(t,x_j) + \beta(t,x_j)\partial_x\zeta(t,x_j))
    \ .
\end{align}
Note that the function $p(t,x)$ can be chosen arbitrarily, as mentioned above.

In summary, we have the spin model that corresponds to the quantum field theory of the Majorana fermion in the curved two-dimensional spacetime~(\ref{generalmetric}) as
\begin{equation}
\begin{split}
    H=-\frac{1}{4\varepsilon} \sum_{j=1}^L\bigg\{&
    \left(\frac{\alpha(t,x_j)}{\gamma(t,x_j)} \cos\zeta(t,x_j) + p(t,x_j)\right)\sigma^x_j \sigma^x_{j+1}\\
    - &\left(\frac{\alpha(t,x_j)}{\gamma(t,x_j)} \cos\zeta(t,x_j) - p(t,x_j)\right)\sigma^y_j \sigma^y_{j+1} \\
    - &\left(\beta(t,x_j) + \frac{\alpha(t,x_j)}{\gamma(t,x_j)} \sin\zeta(t,x_j)\right) \sigma^x_j \sigma^y_{j+1} \\
    + &\left(\beta(t,x_j) - \frac{\alpha(t,x_j)}{\gamma(t,x_j)} \sin\zeta(t,x_j)\right) \sigma^y_j \sigma^x_{j+1}\\
    + &\big[2p(t,x_j) - \varepsilon \big(\partial_x p(t,x_j) + 2m\alpha(t,x_j)\\
    &\hspace{0.3cm} - \partial_t\zeta(t,x_j) - \beta(t,x_j)\partial_x\zeta(t,x_j)\big)\big] \sigma^z_j \bigg\} ,
    \label{IsingforQFT}
\end{split}
\end{equation}
where $p(t,x)$ is a free function. 
From Eqs. (\ref{JW}) and (\ref{Psi_cnt}), we find that 
$\sigma^{x,y}=\mathcal{O}(\sqrt{\varepsilon}\Psi)$ and $\sigma^{z}=\mathcal{O}(1)$ in the continuum limit. Thus, in the above expression, 
we need to leave the terms involving $\varepsilon \sigma^z_j$. This can be rewritten as 
\begin{equation}
\begin{split}
 H=&- \frac{1}{2\varepsilon} \sum_{j=1}^{L}\bigg\{ \frac{\alpha(t,x_j)}{\gamma(t,x_j)}\cos\zeta(t,x_j)
 ( c_{j+1}c_j +c_j^\dagger c_{j+1}^\dagger)\\
& + i\frac{\alpha(t,x_j)}{\gamma(t,x_j)}\sin\zeta(t,x_j)
 ( c_{j+1}c_j - c_j^\dagger c_{j+1}^\dagger)\\
& + p(t,x_j)( c_j^\dagger c_{j+1}+ c_{j+1}^\dagger c_j)
 \\
 &+ i\beta(t,x_j)( c_j^\dagger c_{j+1}- c_{j+1}^\dagger c_j)\\
&+\frac{1}{2}\big[2p(t,x_j) - \varepsilon\big(\partial_x p(t,x_j) + 2m\alpha(t,x_j)\\
&\hspace{0.2cm} - \partial_t\zeta(t,x_j) - \beta(t,x_j)\partial_x\zeta(t,x_j)\big)\big]
(1-2c_j^\dagger c_j)\bigg\} .
 \label{IsingforQFTc}
\end{split}
\end{equation}
in terms of fermionic operators via the Jordan-Wigner transformation. 

The parameters of the QFT in a curved spacetime, $\alpha$, $\beta$, $\gamma$ and $\zeta$, can conversely be determined from the parameters of the spin model, $J^{ab}_j$ and $h_j$, in the following way: 
\begin{equation}
    \begin{aligned}
    m \alpha(t,x_j) =& - 2h_j 
            + 2\mathrm{Re} J^{-+}_j - \frac{1}{2}\mathrm{Re} J^{-+}_{j+1} + \frac{1}{2}\mathrm{Re} J^{-+}_{j-1}  \\
             + \frac{1}{2} &
            \left[\frac{d}{dt}\mathrm{arg} J^{++}_j + \mathrm{Im} J^{-+}_j (\mathrm{arg} J^{++}_{j+1} - \mathrm{arg} J^{++}_{j-1}) \right],\\
        \beta(t,x_j) =& 2\varepsilon \mathrm{Im} J^{-+}_j = - 2\varepsilon \mathrm{Im} J^{+-}_j ,\\
        \gamma (t,x_j) =& \frac{\alpha(t,x_j)}{2\varepsilon|J^{++}_j|} = \frac{\alpha(t,x_j)}{2\varepsilon|J^{--}_j|} ,\\ 
        \zeta (t,x_j) =& \mathrm{arg} J^{++}_j = - \mathrm{arg} J^{--}_j .
    \end{aligned}
    \label{eq:Ising_to_QFT}
\end{equation}
Thus, once a quantum spin model described by the Hamiltonian (\ref{Hgeneral}) is given, we can obtain the corresponding field theory of a Majorana fermion in a curved spacetime. In particular, we can read metric functions representing the curved spacetime.

As explained in section~\ref{sec:Marjorana_2D}, the sign of the mass $m$ is physically irrelevant in the continuum limit. In the spin system, the flip of the sign of $m$ is realized by the unitary transformation $\sigma^x_j  \to (-1)^{j-1} \sigma^y_j$, $\sigma^y_j \to (-1)^{j-1} \sigma^x_j$, $\sigma^z_j \to -\sigma^z_j$ together with the sign inversion of the arbitrary functions $p\to -p$ and $\zeta\to -\zeta$.

\subsection{On the choice of free functions: $p(t,x)$ and $\zeta(t,x)$}

We have two free functions $p(t,x)$ and $\zeta(t,x)$ in Eq.~(\ref{IsingforQFT}) to provide a continuum field theory. 
The free function $\zeta$ corresponds to the rotational degrees of freedom of the spin system around the $z$-axis: $\sigma^\pm_j \to e^{\pm i \zeta_j(t)/2} \sigma^\pm_j$. In this sense, for any function $\zeta$, the Hamiltonian~(\ref{IsingforQFT}) represents the same spin system. 
Similarly, in the field theory side, this functional freedom $\zeta$ corresponds to the freedom to choose the phase of the Majorana field, as seen in Eq.~(\ref{Psidef}).

On the other hand, the free function $p$ cannot be eliminated even by redefining the spin operators.
In this sense, 
if the function $p$ is different, 
the Hamiltonian~(\ref{IsingforQFT}) represents a distinct spin system for a finite $L$.
However, in the continuum limit $L\to\infty$ with fixed $\ell$, 
the Hamiltonian converges to that of the same field theory, irrespective of the function $p$.

One of the simplest choices of these free functions is
\begin{equation}
    p(t,x)=\frac{\alpha(t,x)}{\gamma(t,x)}\ ,\quad \zeta(t,x)=0\ .
\end{equation}
Then, the Hamiltonian~(\ref{IsingforQFT}) reduces to
\begin{equation}
\begin{split}
    H=&-\frac{1}{4\varepsilon} \sum_{j=1}^L\Bigg\{
    2\frac{\alpha(t,x_j)}{\gamma(t,x_j)}\sigma^x_j \sigma^x_{j+1}\\
    &- \beta(t,x_j) (\sigma^x_j \sigma^y_{j+1} -\sigma^y_j \sigma^x_{j+1})+ \bigg[2\frac{\alpha(t,x_j)}{\gamma(t,x_j)}\\
    & - \varepsilon \left(\partial_x \left(\frac{\alpha(t,x_j)}{\gamma(t,x_j)}\right) + 2m\alpha(t,x_j)\right)\bigg] \sigma^z_j \Bigg\} .
    \label{IsingforQFT_simple}
\end{split}
\end{equation}
This spin model corresponds to the transverse-field Ising model with ``transverse'' Dzyaloshinskii-Moriya interaction (TI-DM model). Another simple choice is
\begin{equation}
    p(t,x)=0\ ,\quad \zeta(t,x)=\pi/2\ ,
\end{equation}
which leads to
\begin{equation}
\begin{split}
    H=\frac{1}{4\varepsilon}& \sum_{j=1}^L\bigg\{
     \frac{\alpha(t,x_j)}{\gamma(t,x_j)}  \left(\sigma^x_j \sigma^y_{j+1} +  \sigma^y_j \sigma^x_{j+1}\right)\\
    +&  \beta(t,x_j) (\sigma^x_j \sigma^y_{j+1} -\sigma^y_j \sigma^x_{j+1})+2 \varepsilon  m\alpha(t,x_j) \sigma^z_j \bigg\}.
    \label{IsingforQFT_simple2}
\end{split}
\end{equation}
The first term in parentheses is called the ``$\Gamma$ term'' in spin language. 

These choices represent a minimal spin model that can describe the QFT of Majorana fermions in any $(1+1)$-dimensional spacetime. Of course, one can choose other functions if they are more convenient for actual simulations.

Note that for a specific choice of $p(t,x)$, such as $p = 0$ in the case of the Minkowski metric with $\alpha = \gamma = 1$ and $\beta = 0$, additional fermionic modes can appear at momenta away from $k \sim 0$ in the low-energy regime, reminiscent of the fermion doubling problem identified by Kenneth Wilson. This implies that the function $p(t,x)$ can be used to control and suppress such doubling, effectively playing the same role as Wilson's term in the lattice gauge theory~\cite{Wilson:1974sk,Wilson:1975id,wilson_1977}. Remarkably, in our dictionary, this term is not inserted by hand but emerges naturally from the structure of the mapping itself.

\section{Simulation of particle production in expanding universe in the Ising model}
\label{sec:particle_production}

As an example which demonstrates that the spin systems work as simulators of the quantum field theories in curved spacetimes, we consider an expanding universe: 
\begin{equation}
 ds^2=-dt^2+a^2(t)dx^2 = a^2(\eta)(-d\eta^2+dx^2)\ .
\end{equation}
This is called the Friedmann-Lema\^{\i}tre-Robertson-Walker (FLRW) metric. The time coordinates $t$ and $\eta\equiv \int dt/a(t)$ are called the cosmological time and conformal time, respectively. The function $a(t)$ is referred to as the scale factor. 
In Sec.~\ref{subsec:Majorana_form}, we focus on a continuum theory of the Majorana fermion and discuss the particle production in the expanding universe. In Sec.~\ref{subsec:Ising_form}, we will simulate the particle production using the discrete theory of the spin system.
Note that by using another quantum system the quantum simulation of the particle production for Dirac fermions was also discussed in~\cite{Fulgado-Claudio:2022fut}, for example.

\subsection{Continuum theory}
\label{subsec:Majorana_form}

We will assume that the space is compactified as $x\sim x+ \ell$ and the anti-periodic boundary condition $\Psi(x+\ell)=- \Psi(x)$ is imposed on the Majorana fermion. (We will see that this choice of boundary condition is convenient for comparison  with the spin system.)

We use the conformal time $\eta$ to describe the time evolution of the quantum system. In terms of the general metric~(\ref{generalmetric}), setting $\alpha=\gamma=a(\eta)$ and $\beta = 0$ provides the FLRW metric in the conformal time.
From Eq.~(\ref{MajH}), the Hamiltonian density of the Majorana fermion is 
\begin{equation}
\mathcal{H}=-\frac{1}{2}(\Psi^\dagger \partial_x \Psi^\dagger-\Psi \partial_x \Psi) 
-m a(\eta) \Psi^\dagger \Psi \ ,
\label{MajRW}
\end{equation}
where we have taken $\zeta = 0$ for simplicity.
Thus, a field theory for the Majorana spinor field in the FLRW spacetime is equivalent to that in the flat spacetime with a time-dependent mass. This is similar to cases of a scalar field in the FLRW spacetime~\cite{BD}.
The Hamiltonian density~(\ref{MajRW}) yields the Heisenberg equations as 
\begin{equation}
\begin{split}
&i\frac{\partial}{\partial \eta}\Psi +\frac{\partial}{\partial x}\Psi^\dagger + ma(\eta)\Psi= 0\ ,\\ &i\frac{\partial}{\partial \eta}\Psi^\dagger-\frac{\partial}{\partial x}\Psi -m a(\eta) \Psi^\dagger= 0\ .
\end{split}
\end{equation}
We apply the Fourier transform as
\begin{equation}
\begin{split}
    \Psi(\eta,x)&=\frac{1}{\sqrt{\ell}}\sum_{k\in K} e^{ikx} \Psi_k(\eta)\ ,\\ 
    \Psi^\dagger (\eta,x)&=\frac{1}{\sqrt{\ell}}\sum_{k\in K} e^{ikx} \Psi^\dagger_{-k}(\eta)\ ,
\end{split}
\end{equation}
where the domain of the wave number is
\begin{equation}
K=\left\{\frac{2\pi}{\ell}\left(n-\frac{1}{2}\right)\bigg|n\in \bm{Z}\right\}\ .
\label{Kdef_cont}
\end{equation}
After this Fourier transformation, the Heisenberg equations are rewritten as 
\begin{equation}
\left(
i\frac{d}{d\eta}
-
M_k(\eta)
\right)
\vec{\Psi}_k
=0\ ,\quad
\vec{\Psi}_k\equiv 
\begin{pmatrix}
\Psi_k(\eta) \\
\Psi_{-k}^\dagger(\eta)
\end{pmatrix}
,
\end{equation}
where we define the Hermitian matrix as
\begin{equation}
M_k(\eta)=
    \begin{pmatrix}
-ma(\eta) & -ik \\
ik & ma(\eta)
\end{pmatrix}\ .
\label{Mcont}
\end{equation}
The general solution of the above equation is given by 
\begin{equation}
    \vec{\Psi}_k(\eta)= \gamma_k \vec{\phi}_k(\eta) + \gamma'_k \vec{\phi}'_k(\eta)\ ,
\label{PsiMdef}
\end{equation}
where $\gamma_k$ and $\gamma_{k}'$ represent time-independent operators. 
$\vec{\phi}_k(\eta)$ and $\vec{\phi}'_k(\eta)$ are linearly independent solution of 
\begin{equation}
    \left(i\frac{d}{d\eta}-M_k(\eta)\right)\vec{\phi}_k=0\ ,
    \label{phieq}
\end{equation}
whose components are $c$-numbers. 
Since the matrix $M_k$ is Hermitian, time evolution of $\vec{\phi}_k$ and $\vec{\phi}_k'$ is given by a unitary transformation. 
Thus, the ordinary inner product
\begin{equation}
    (\vec{\phi}_k,\vec{\phi}'_k)=(\vec{\phi}_k)^\dagger \vec{\phi}'_k\ ,
\label{innerprod}
\end{equation}
is time-independent.

The matrix $M_k(\eta)$ defined in Eq.~(\ref{Mcont}) satisfies $\sigma^x M_k(\eta)\sigma^x=-M_{-k}^\ast(\eta)$.
It follows that, once we obtain a solution of Eq.~(\ref{phieq}) for any $k\in K$, we can also generate the other linearly independent solution as 
\begin{equation}
    \vec{\phi}_k' = \sigma^x \vec{\phi}_{-k}^\ast\ .
    \label{phiprim}
\end{equation}
From Eq.~(\ref{PsiMdef}), we can see that components of $\vec{\Psi}_k$ are related as $\vec{\Psi}_k = \sigma^x (\vec{\Psi}_{-k}^\dagger)^T$. Thus, when we choose $\vec{\phi}_k'$ as in Eq.~(\ref{phiprim}), 
we have $\gamma'_k=\gamma_{-k}^\dagger$. Eventually, the Majorana field is written as 
\begin{equation}
    \vec{\Psi}_k(\eta)= \gamma_k \vec{\phi}_k(\eta) + \gamma_{-k}^\dagger \sigma^x \vec{\phi}_{-k}^\ast(\eta)\ .
\end{equation}
If we choose the mode functions $\vec{\phi}_k$ so that $\vec{\phi}_k$ and $\vec{\phi}'_k=\sigma^x \vec{\phi}_{-k}^\ast$ are orthonormal: 
\begin{equation}
    (\vec{\phi}_k,\vec{\phi}_k)=(\sigma^x \vec{\phi}_{-k}^\ast,\sigma^x \vec{\phi}_{-k}^\ast)=1\ ,\quad 
(\vec{\phi}_k,\sigma^x \vec{\phi}_{-k}^\ast)=0\ ,
\label{orth}
\end{equation}
then $\gamma_k$ satisfies the canonical anti-commutation relation $\{\gamma_k,\gamma_{k'}^\dagger\}=\delta_{kk'}$.

The choice of such mode functions $\vec{\phi}_k$ is not unique. Suppose that there is another choice of mode functions $\vec{\Phi}_k$ and the Majorana field is expanded into 
\begin{equation}
    \vec{\Psi}_k(\eta)= \Gamma_k \vec{\Phi}_k(\eta) + \Gamma_{-k}^\dagger \sigma^x \vec{\Phi}_{-k}^\ast(\eta)\ .
\end{equation}
Again we have assumed that $\vec{\Phi}_k$ and $\sigma^x \vec{\Phi}_{-k}^\ast$ are orthonormal. Then, we can easily obtain the relation between $\gamma_k$ and $\Gamma_k$ as
\begin{equation}
 \Gamma_k=(\vec{\Phi}_k, \vec{\phi}_k) \gamma_k + (\vec{\Phi}_k, \sigma^x \vec{\phi}_{-k}^\ast) \gamma_{-k}^\dagger \ ,
\end{equation}
which gives the Bogoliubov transformation between $\gamma_k$ and $\Gamma_k$.
We define the ``vacuum'' state $|\Omega\rangle$ in terms of $\gamma_k$, such that $\gamma_k|\Omega\rangle=0$ for any $k\in K$. Then, the expectation value of the particle number in terms of $\Gamma_k$ is computed as 
\begin{equation}
 n_k\equiv \langle \Omega| \Gamma_k^\dagger \Gamma_k |\Omega\rangle=
|(\vec{\Phi}_k, \sigma^x \vec{\phi}_{-k}^\ast)|^2\ .
\label{Num1}
\end{equation}
This is time-independent because of the conservation of the inner product defined by Eq.~(\ref{innerprod}).

\subsection{Discrete theory: Transverse-field Ising model}
\label{subsec:Ising_form}

In the previous subsection, we have set $\alpha=\gamma=a(\eta)$ and $\beta = \zeta = 0$ in the Hamiltonian density~(\ref{MajH}) so that we obtain the field theory for the Majorana fermion in the FLRW metric with the conformal time.
Thus, from Eq.~(\ref{IsingforQFT}), the Ising model which corresponds to the Majorana fermion in the expanding universe is given by
\begin{equation}
    H=-\frac{1}{2\varepsilon} \sum_{j=1}^L\bigg[
    \sigma^x_j \sigma^x_{j+1} 
    +(1-\varepsilon m a(\eta)) \sigma^z_j  
    \bigg]\ ,
    \label{HRWdisc}
\end{equation}
where we have chosen the free function as $p(\eta ,x) = 1$. 
The above Hamiltonian is known as the transverse-field Ising model. The transverse magnetic field depends on time $\eta$. After the Jordan-Wigner transformation, from Eq.~(\ref{IsingforQFTc}), the above Hamiltonian is written as
\begin{equation}
\begin{split}
 H=- \frac{1}{2\varepsilon} \sum_{j=1}^{L}&\bigg[c_{j+1}c_j +c_j^\dagger c_{j+1}^\dagger
 +c_j^\dagger c_{j+1}+ c_{j+1}^\dagger c_j
 \\&+(1-\varepsilon m a(\eta) ) (1-2c_j^\dagger c_j)\bigg]\ .
\end{split}
\end{equation}
We apply the Fourier transformation of the operator $c_j$ as 
\begin{equation}
 c_j = \frac{1}{\sqrt{L}}\sum_{\kappa \in \mathcal{K}} e^{i \kappa j} c_\kappa\ .
\end{equation}
For the Hilbert space $\mathcal{H}_\textrm{even}$ and $\mathcal{H}_\textrm{odd}$, 
the Jordan-Winger fermion operator satisfies $c_L=-c_1$ (anti-periodic) and $c_L=c_1$ (periodic), respectively.
When the transverse-magnetic field is constant (i.e., $a(\eta)=\text{const.}$), it is known that the ground state is in $\mathcal{H}_\textrm{even}$.
We assume that $a(\eta)\to \textrm{const.}$ as $\eta\to -\infty$ and the quantum state starts from the ground state initially. Then, by the conservation of the oddity of the total fermion number, we can always restrict our attention to $\mathcal{H}_\textrm{even}$. 
Thus, hereafter, we will focus only on the anti-periodic boundary condition. Then, the domain of the wave number $\kappa$ is given by
\begin{equation}
\mathcal{K}=\left\{\frac{2\pi}{L} \left(n-\frac{1}{2}\right)  \bigg| n=-\frac{L}{2}+1,\cdots,\frac{L}{2}\right\}\ .
\label{Kdef_disc}
\end{equation}
In the continuum theory, we have also introduced the wave number $k \in K$ as in Eq.~(\ref{Kdef_cont}). 
The relation between the wave numbers in the continuum  and discrete theories is given by 
\begin{equation}
    k=\frac{\kappa}{\varepsilon}\ ,
\end{equation}
while $\kappa$ should be bounded due to the discreteness of the space, which corresponds to an ultraviolet cutoff $\sim 1/\varepsilon$.
Note that we have taken the long-wavelength limit~(\ref{lowk}) when we consider the continuum limit in the Ising model. In the discrete theory,  the long-wavelength  limit is written as
\begin{equation}
    \kappa \ll 1 \Leftrightarrow n\ll L\ .
\end{equation}
We can only think about the wave number satisfying the above condition to simulate the quantum field theory by the Ising model.

The Heisenberg equation in the momentum space is given by
\begin{equation}
    \left(i\frac{d}{d\eta}-M_\kappa (\eta)\right) \vec{c}_\kappa = 0 \ ,\quad \vec{c}_\kappa\equiv \mtx{c_\kappa(\eta) \\ c_{-\kappa}^\dagger(\eta)}\ ,
    \label{Heqdisc}
\end{equation}
where 
\begin{equation}
\begin{split}
    M_\kappa (\eta)&=\\ \frac{1}{\varepsilon} &\mtx{1-\cos \kappa-\varepsilon m a(\eta)  & -i\sin \kappa \\ i\sin \kappa & -(1-\cos \kappa-\varepsilon m a(\eta)) }.
    \label{Mdisc}    
\end{split}
\end{equation}
In a manner similar to section~\ref{subsec:Majorana_form}, the solution of the Heisenberg equation is written as
\begin{equation}
    \vec{c}_\kappa(\eta)= \gamma_\kappa \vec{\phi}_\kappa(\eta) + \gamma_{-\kappa}^\dagger \sigma^x \vec{\phi}\,{}^\ast_{-\kappa}(\eta)\ ,
\end{equation}
where  $\vec{\phi}_\kappa$ is a solution of $(id/d\eta-M_\kappa (\eta))\vec{\phi}_\kappa=0$.
We choose $\vec{\phi}_\kappa$ so that $\vec{\phi}_\kappa$ and $\sigma^x \vec{\phi}\,{}^\ast_{-\kappa}$ are orthonormal with respect to the inner product $(\vec{\phi},\vec{\phi}')=\vec{\phi}^\dagger \vec{\phi}'$.

Suppose that the other mode functions $\vec{\Phi}_\kappa$ exist and $\vec{c}_\kappa$ are expanded in terms of $\vec{\Phi}_\kappa$ into  
\begin{equation}
    \vec{c}_k(\eta)= \Gamma_\kappa \vec{\Phi}_\kappa(\eta) + \Gamma_{-\kappa}^\dagger \sigma^x \vec{\Phi}_{-\kappa}^\ast(\eta)\ .
\end{equation}
Here, $\vec{\Phi}_k$ and $\sigma^x \vec{\Phi}_{-k}^\ast$ are orthonormal. 
Defining the ``vacuum'' state $|\Omega\rangle$ such that $\gamma_\kappa|\Omega\rangle=0$ for any $\kappa\in \mathcal{K}$, we obtain the expectation value of the particle number in terms of $\Gamma_\kappa$ is computed as 
\begin{equation}
 n_\kappa\equiv \langle \Omega| \Gamma_\kappa^\dagger \Gamma_\kappa |\Omega\rangle=
|(\vec{\Phi}_\kappa, \sigma^x \vec{\phi}_{-\kappa}^\ast)|^2\ .
\label{Num2}
\end{equation}

\subsection{Exactly solvable example}

\begin{figure}[t]
  \centering
\subfigure[$\Delta\eta=1$]
 {\includegraphics[scale=0.35]{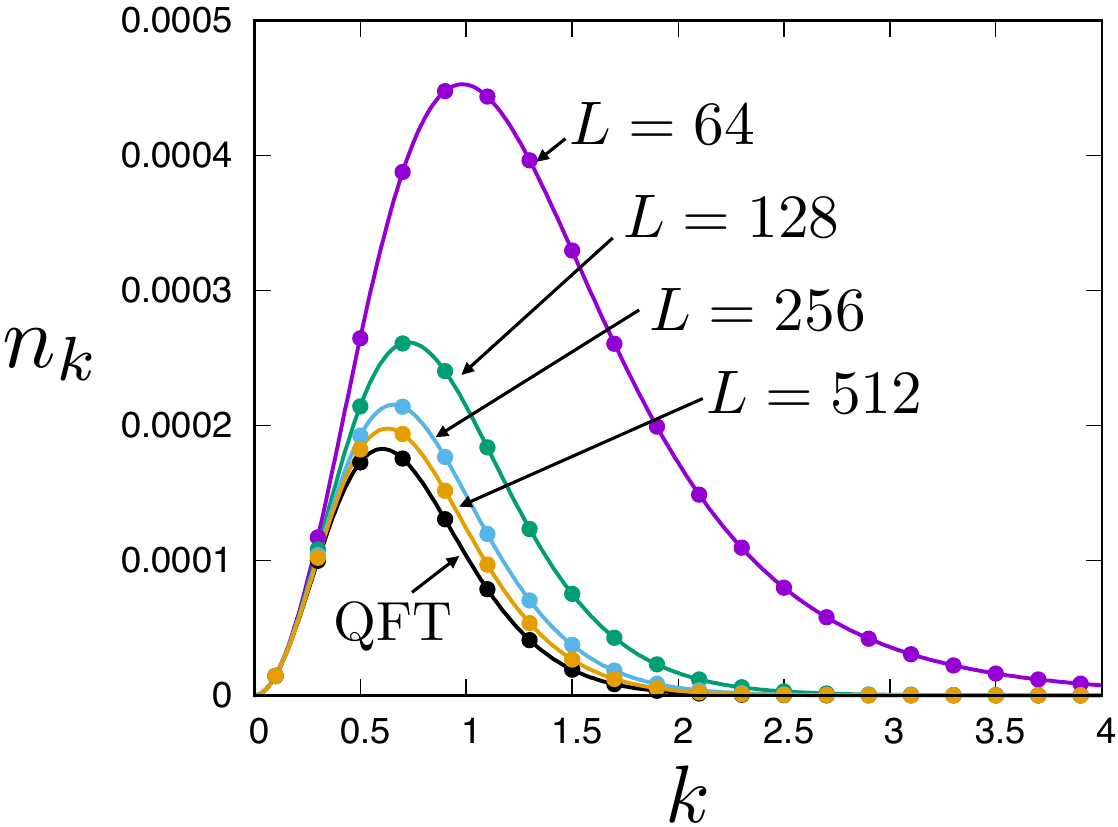}
  }
  \subfigure[$\Delta\eta=0.2$]
 {\includegraphics[scale=0.35]{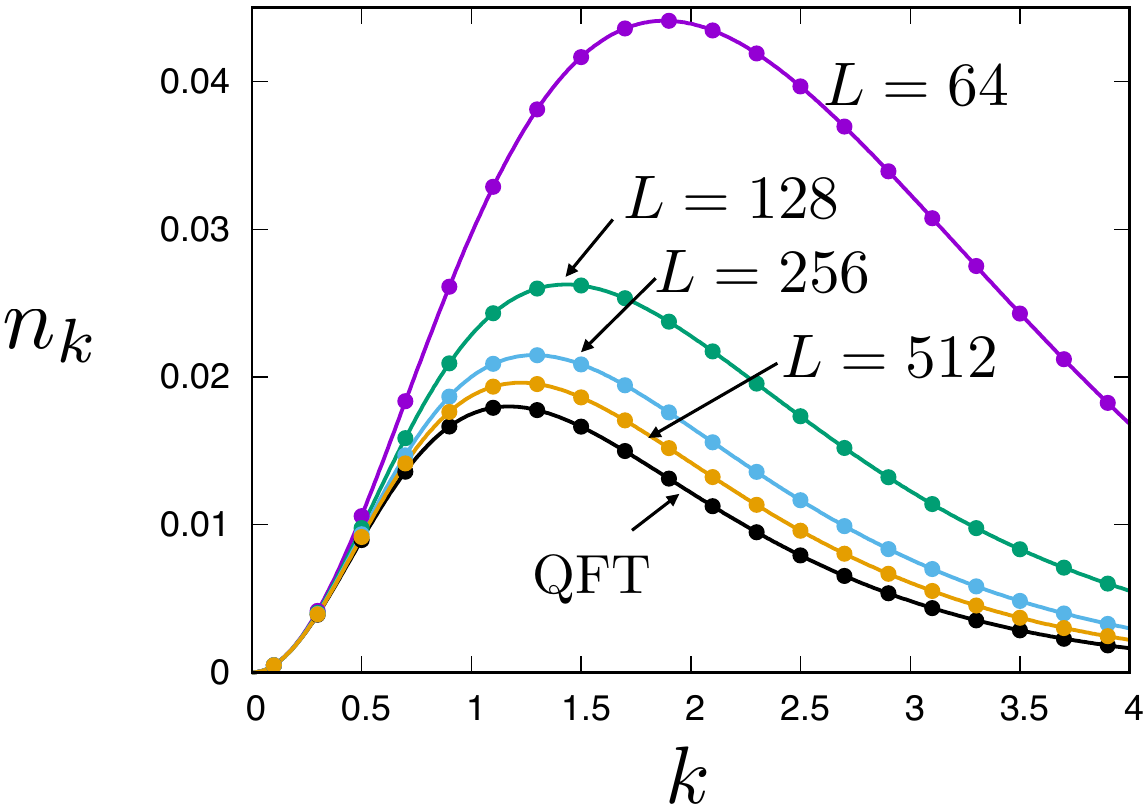}
  }
 \caption{The number of particles produced by the cosmic expansion as the function of the wave number $k=\kappa/\varepsilon$. Parameters are $m=a_1=1$, $a_2=2$ and $\ell/2\pi=5$. The upper and lower panels are for $\Delta\eta=1$ and $0.2$, respectively.
For discrete theory, the number of sites are varied as $L=64,128,256,512$.
}
 \label{spectrum}
\end{figure}

\begin{figure}[t]
\begin{center}
\includegraphics[scale=0.3]{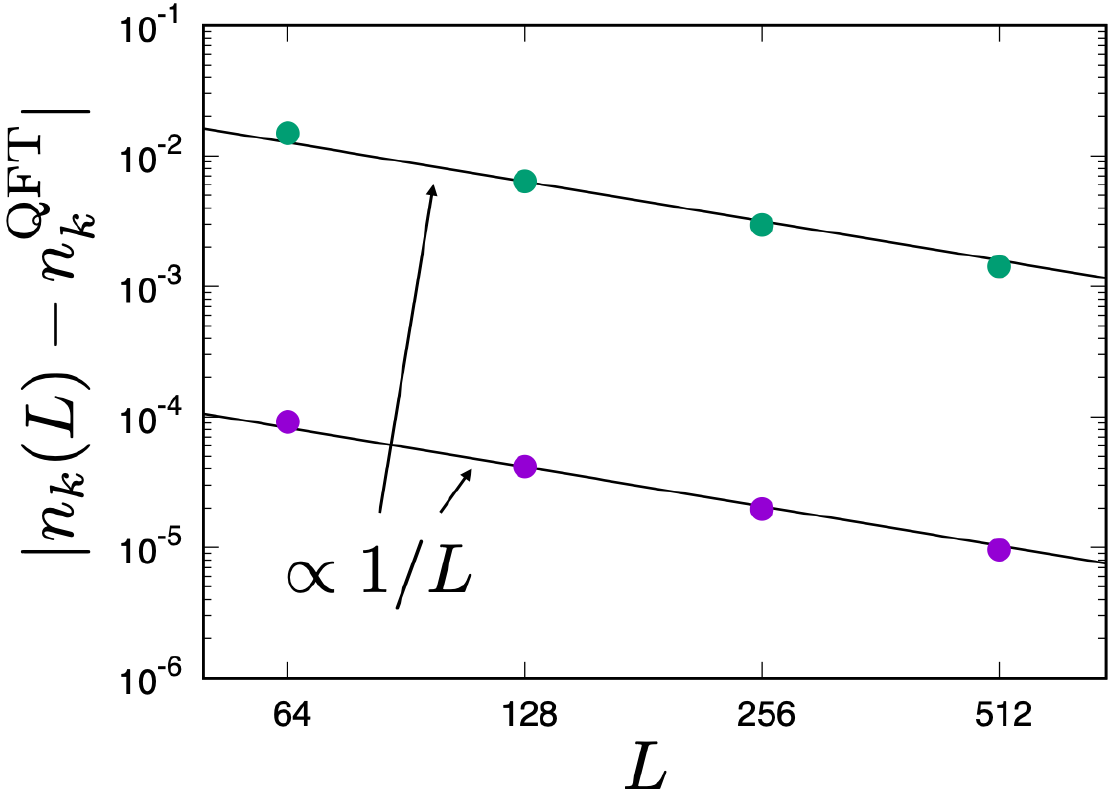}
\end{center}
\caption{The difference between the number of particles produced in discrete and continuum theories, $|n_k(L)-n_k^\text{QFT}|$, for given typical wave numbers. The purple and green points correspond to the data in Fig.~\ref{spectrum} (a) and (b), respectively. The wave number is fixed as (a) $k=0.5$ and  (b) $k=1.1$. 
}
\label{convergence}
\end{figure}

We consider an expanding universe represented by the following scale factor:
\begin{equation}
 a(\eta)=\frac{a_2+a_1}{2}+\frac{a_2-a_1}{2}\tanh \left(\frac{\eta}{\Delta \eta}\right) \ .
 \label{tanh}
\end{equation}
This satisfies $a(\eta)\to a_1$ ($\eta\to-\infty$) and $a(\eta)\to a_2$ ($\eta\to\infty$). 
In this case, we have analytical solutions of mode functions both in continuum and discrete theories. 
The detailed calculation is summarized in appendix~\ref{detail}. 
Once we have two sets of mode functions, we can compute the number of produced particles for a given wave number using Eqs.~(\ref{Num1}) and (\ref{Num2}).
As the quantum state $|\Omega\rangle$, we take the ground state at the sufficiently early time. 
Figure~\ref{spectrum} shows the number of particles produced by the cosmic expansion with $m=a_1=1$, $a_2=2$ and $\ell/2\pi =5$. 
The upper and lower panels are for $\Delta\eta = 1$ and $\Delta \eta=0.2$.
Both results for the continuum and discrete theories are shown together. For the discrete theory, we take $k=\kappa/\varepsilon$ as the horizontal axis for the visibility of the continuum limit $L\to\infty$. The number of sites is varied as $L=64, 128, 256, 512$. 
Since we consider that $\ell$ is fixed, the lattice spacing $\varepsilon=\ell/L$ depends on $L$ while the wave numbers $k$ are defined at the same discrete points for any $L$.
In Fig.~\ref{spectrum}, we plot $n_k$ as a function of the wave number at the discrete points represented by dots, together with a continuum variable for visibility.
We find that the spectrum tends to converge to that of the continuum theory as $L$ increases. 
Indeed, the difference between the number of particles produced in discrete and continuum theories, $|n_k(L)-n_k^\text{QFT}|$, depends on $O(1/L)$, as shown in Fig.~\ref{convergence}. 
Especially, for $L\gtrsim 100$, the spectrum qualitatively agrees with that of the continuum theory. This result demonstrates that the spin model can work as a simulator of QFT in the curved spacetime. 
As the timescale of the cosmic expansion $\Delta \eta$ decreases, more particles are produced. 
Also, typical wave number of produced particles 
 increases as $\Delta \eta$ decreases.
 Now, a physical length scale of the systems is given by $\ell$.
 A typical Hubble scale is $H_\eta \equiv a^{-2} da/d\eta \approx 1/a\Delta\eta$ during the cosmic expansion.
 Therefore, the present phenomena should be governed by the relation between the values of two dimensionless parameters, the mass scale $m\ell$ and the Hubble scale $H_\eta \ell$.

In Appendix~\ref{on_exp_measure}, we provide a relation between the number of produced particles $n_k$ and the correlation functions of spin operators, which will be needed when we consider experiments to observe particle production in the expanding universe through the spin system.

\section{Comments on the Unruh effect in the spin system}
\label{sec:unruh}

\subsection{Entanglement Hamiltonian of the transverse-field Ising model}

Let us consider a spin system and decompose the system into $A$ and $B$. For a given state $|\psi \rangle$ in the spin system, the entanglement Hamiltonian $K_A$ is defined as 
\begin{equation}
 \textrm{Tr}_B |\psi \rangle \langle  \psi | \propto e^{-K_A}\ ,
\end{equation}
where $\text{Tr}_B$ represents the partial trace of in the region $B$. (See \cite{Dalmonte:2022rlo} for a nice review of the entanglement Hamiltonian.) 
In general, the entanglement Hamiltonian is non-local and has a complicated structure.
However, in some special cases, we obtain its explicit expression. One example is the transverse-field Ising model with infinite sites:
\begin{equation}
    H=-\sum_{j=-\infty}^\infty (J \sigma^x_j \sigma^x_{j+1} + h\sigma^z_j)\ .
    \label{eq:infinite_TFI}
\end{equation}
We will focus only on the ordered phase $J>h$. 
Taking $A=\{j\in \bm{Z}|j> 0\}$ and $B=\{j\in \bm{Z}|j\leq 0\}$, we have explicitly the expression for the entanglement Hamiltonian with respect to the ground state as~\cite{Davies:1988zz,Truong}
\begin{equation}
    K_A = -2I(r')\sum_{j=1}^\infty \left[j\sigma^x_j \sigma^x_{j+1} + r \left(j-\frac{1}{2}\right)\sigma_j^z\right]\ ,
    \label{EH}
\end{equation}
where $r=h/J$, 
$r'=\sqrt{1-r^2}$ and $I(r')$ is the complete elliptic integral of the first kind. 

\subsection{Unruh effect in quantum field theory}

A uniformly accelerating observer in the Minkowski spacetime will perceive a thermal bath. The temperature of the thermal bath is given by $T_U\equiv\kappa/(2\pi)$, where $\kappa$ is the proper acceleration of the observer.
This is known as the Unruh effect (or Fulling–Davies–Unruh effect)~\cite{Fulling:1972md,Davies:1974th,Unruh:1976db}.
The mathematical expression for the Unruh effect is simply given by
\begin{equation}
 \textrm{Tr}_B |\Omega \rangle \langle \Omega | \propto \exp(-H_R/T_U)\ ,
\end{equation}
where $|\Omega \rangle$ is the vacuum state of QFT in the Minkowski spacetime, $ds^2 = -dT^2 + dX^2$. 
(Here, we focus only on the two-dimensional QFT.)
We have decomposed the spatial coordinate $X$ at a time slice $T=0$ into the right and left regions as $A=(0,\infty)$ and $B=(-\infty,0)$.
The operator $H_R$ is the Hamiltonian of QFT in the Rindler spacetime:
\begin{equation}
 ds^2= -\kappa^2 x^2 dt^2 + dx^2\ , \label{eq:Rindler}
\end{equation}
where the coordinate transformation is given by $T=x\sinh(\kappa t)$ and $X=x\cosh(\kappa t)$.
Note that $H_R$ corresponds to a generator of time translation with respect to the proper time of the uniformly accelerating observer at $x=1/\kappa$.

\subsection{Confirmation of the Unruh effect through the spin system}

The open spin system corresponding to QFT in the Minkowski spacetime is given by
\begin{equation}
    H=-\frac{1}{2\varepsilon} \bigg[\sum_{j=-L/2+1}^{L/2-1}
    \sigma^x_j \sigma^x_{j+1} 
    +\sum_{j=-L/2+1}^{L/2} (1-\varepsilon m) \sigma^z_j  
    \bigg]\ ,
\end{equation}
where we have sifted the origin of the site index so that $-L/2+1\leq j \leq L/2$.
The lattice spacing is $\varepsilon = \ell/L$ and the spatial coordinate of the $j$-th site is $X_j=\varepsilon j$. 
If we take the limit $L\to\infty$ for a fixed $\varepsilon$ (i.e., $\ell\to\infty$), we have the Hamiltonian of the transverse-field Ising model with infinite sites (\ref{eq:infinite_TFI}).
Then, the continuum limit $\varepsilon \to 0$ corresponding to the limit $J = 1/(2\varepsilon) \to \infty$ provides a QFT on the whole spatial region $-\infty < X < \infty$ in the Minkowski spacetime.
Note that the mass, which is given by $m=2J-2h=2J(1-r)$, should remain finite.

Now, from Eq.~(\ref{EH}), we consider the entanglement Hamiltonian multiplied by the Unruh temperature as 
\begin{equation}
\begin{split}
    T_U K_A =& -2I(r') T_U \bigg[\sum_{j=1}^{L/2-1}
    j \sigma^x_j \sigma^x_{j+1} 
    +\sum_{j=1}^{L/2} r \left(j-\frac{1}{2}\right) \sigma^z_j  
    \bigg]  \\ &\quad (L\to \infty) .
\end{split}
\end{equation}
The parameters of the spin model (\ref{Hgeneral}) are given by 
\begin{equation}
\begin{split}
    J^{++}_j &= J^{--}_j = J^{+-}_j = J^{-+}_j = 2I(r') T_U j ,\\
    h_j &= 2I(r')T_U r\left(j - \frac{1}{2}\right) .
\end{split}
\end{equation}
By using ($\ref{eq:Ising_to_QFT}$), we can obtain the corresponding parameters of the QFT as follows:
\begin{equation}
    \begin{aligned}
        \alpha_j &= 4I(r')T_U \varepsilon \left(j - \frac{1}{2}\right) ,\\
        \gamma_j &= 1 - \frac{1}{2j} ,\quad\beta_j = \zeta_j = 0 .
    \end{aligned}
\end{equation}
Introducing $x_j \equiv \varepsilon j = X_j$ and taking the continuum limit $\varepsilon = 1/2J \to 0$, we find  
\begin{equation}
    \alpha = 2\pi T_U x ,\quad  \gamma = 1 ,\quad  \beta = \zeta = 0 ,
\end{equation}
where we have used $I(r')=(\pi/2)(1+r'{}^2/4+\cdots)=(\pi/2)(1+\varepsilon m/2+\cdots)$.
These are the metric functions of the Rindler spacetime with $\kappa=2\pi T_U$, as shown in (\ref{eq:Rindler}).
Thus, the entanglement Hamiltonian~(\ref{EH}) is directly related to the Rindler Hamiltonian $H_R$ in the continuum limit and exhibits the Unruh effect.
Note that a related discussion on the XXZ chain has been presented in~\cite{Okunishi:2019dmv}.

\section{Summary and discussion}

In this paper, we examined the mapping from the spin systems to the model with the free Majorana field in the $(1+1)$-dimesional curved background by taking the proper limit. 
In the present model, we consider the spin model with general XY-exchange, Z-component Dzyaloshinskii-Moriya, and Z-component $\Gamma$ interactions (XY-DM-$\Gamma$ model).  
By taking the Jordan-Wigner transformation, we obtain the corresponding fermionic model and find that one can obtain the model with Majorana field in the curved background in the continuum limit if we properly choose the time dependence and spatial dependence of parameters in the spin model. 

To demonstrate that the spin systems work as simulators of the quantum field theories in curved spacetimes, we employ the model where the analytical solution is available in both the continuum and discrete theories. 
As a concrete example, we consider the particle production in the expanding universe. 
The model consists of the free Majorana field with time dependent scale factor. 
We found that the number of produced particle calculated in the corresponding spin model tends to converge to that for the continuum theory  as the number of sites increased. We also consider the Unruh effect in the basis of our mapping. 
As a consequence, we found the direct connection between the entanglement Hamiltonian in the spin system and the Rindler Hamiltonian in the continuum theory. 

We comment on the experimental realization of our protocol to simulate the quantum field theory in the curved background. 
The Ising models are widely used as test-beds due to its simplicity. 
For instance, the quantity invented recently, so-called out-of-time-ordered correlator, is observed in the Ising system~\cite{Nie:2019pxk}. 
The experimental realization of the Ising model is also achieved in the optical lattice system, where the parameters are tunable~\cite{Simon:2011}. 
In the optical lattice system, the entanglement entropy is also measured~\cite{Islam:2015mom}. Later the protocol to measure the entanglement Hamiltonian is also proposed~\cite{Kokail:2020opl}. 
Since the model presented in this paper is similar to the Ising model, and because parameter modulation is feasible in certain experimental systems, along with existing methods for measuring key quantities such as the entanglement Hamiltonian, our protocol is expected to effectively reveal the physics of field theory in a curved background. Moreover, more advanced programmable quantum systems, such as superconducting quantum processors~\cite{Devoret2013-bg,Wendin2017-io}, hold significant potential to provide a more flexible platform for studying the dynamics of more general spin systems ~\eqref{IsingforQFT} that mimic the curved spacetime physics of interest.

In this work, we established a complete dictionary to map field theory in curved backgrounds to spin systems in $(1+1)$ dimensions. Since our framework has been shown to work well, a future direction could be its application to problems that remain unsolved in the continuum limit or are difficult to study experimentally. One possibility of interest is the observation of thermal radiation in black hole spacetimes or the inflationary universe.
The generalization of our method to interacting fields would also be an interesting direction for future research. In the non-perturbative regime, theoretical calculations of QFT in curved spacetime become complex. Spin systems could provide an experimental approach to studying such unexplored areas in theoretical physics. 
(For example, the dynamics of self-interacting fermionic fields in an expanding universe has recently been discussed with ultracold atoms in optical lattices~\cite{Fulgado-Claudio:2024xvk}.)

Finally, we would like to emphasize that this approach offers new insights into both QFT in curved spacetimes and quantum many-body spin systems, highlighting the deep interconnections between these domains. Consequently, it is expected that this framework will also pave the way for discovering nontrivial phenomena in quantum spin systems inspired by the principles of curved spacetime physics.

\begin{acknowledgments}
We would like to thank Tadashi Kadowaki, Masahiro Nozaki, Yohichi Suzuki, and Etsuko Itou for useful discussions and comments.
The work of K.\ M.\ was supported in part by JSPS KAKENHI Grant Nos.\ JP20K03976, JP21H05186 and JP22H01217.
The work of D.\ Y.\ was supported by JSPS KAKENHI Grant Nos.~21H05185, 23K22442, 23K25830, 24K06890, and JST PRESTO Grant Nos.~JPMJPR2118 and JPMJPR245D. 
The work of R.\ Y.\ was supported by JSPS KAKENHI Nos.~19K14616, 20H01838, and 25K07156. 
\end{acknowledgments}

\appendix

\section{Canonical quantization of Majorana fermions and a choice of local Lorentz frame}
\label{CanQ}

\subsection{Canonical quantization of Majorana fermions}

From the Lagrangian density~(\ref{PsiLag}), conjugate momenta of $\Psi$ and $\Psi^\dagger$ are obtained as follows: 
\begin{equation}
        \pi_\Psi=\mathcal{L}\frac{\overset{\leftarrow}{\partial}}{\partial(\partial_t\Psi)}=\frac{i}{2} \Psi^\dagger\ ,\quad 
        \pi_{\Psi^\dagger}=\mathcal{L}\frac{\overset{\leftarrow}{\partial}}{\partial(\partial_t\Psi^\dagger)}=\frac{i}{2} \Psi \ ,
        \label{CanoMom}
\end{equation}
where we have defined the conjugate momenta by right derivatives. Then, Hamiltonian density $\mathcal{H}=\pi_\Psi \partial_t\Psi + \pi_{\Psi^\dagger}\partial_t\Psi^\dagger - \mathcal{L}$ becomes Eq.~(\ref{MajH}). 
Since Eq.~(\ref{CanoMom}) cannot be solved in terms of $\partial_t\Psi$ and $\partial_t \Psi^\dagger$, the current system is singular and 
Eq.~(\ref{CanoMom}) should be regarded as constraints. We define constrains $\phi_1$ and $\phi_2$ as
\begin{equation}
\phi_1=\pi_{\Psi}-\frac{i}{2}\Psi^\dagger \ ,\quad 
    \phi_2=\pi_{\Psi^\dagger} - \frac{i}{2} \Psi  .
\end{equation}
The Poisson brackets among these constraints are 
\begin{equation}
\begin{split}
    &\{\phi_1(x),\phi_1(y)\}_P=\{\phi_2(x),\phi_2(y)\}_P=0\ ,\\
    &\{\phi_1(x),\phi_2(y)\}_P=\{\phi_2(x),\phi_1(y)\}_P=-i \delta (x-y)\ .
    \end{split}
\end{equation}
Here, we always consider the equal-time Poisson brackets and suppress the argument $t$.
The other Poisson brackets, also, are 
\begin{equation}
\begin{split}
    &\{\Psi(x),\phi_1(y)\}_P=\delta(x-y)\ ,\quad \{\Psi(x),\phi_2(y)\}_P=0\ ,\\
    &\{\Psi^\dagger(x),\phi_1(y)\}_P=0\ ,\quad \{\Psi^\dagger(x),\phi_2(y)\}_P=\delta(x-y)\ .
\end{split}
\end{equation}
We define the matrix $C(x,y)$ as
\begin{equation}
\begin{split}
    C(x,y)&=\mtx{\{\phi_1(x),\phi_1(y)\}_P & \{\phi_1(x),\phi_2(y)\}_P \\ \{\phi_2(x),\phi_1(y)\}_P & \{\phi_2(x),\phi_2(y)\}_P}\\
    &=\mtx{0 & -i  \\ -i & 0}\delta(x-y)\ ,
\end{split}
\end{equation}
and also its ``inverse matrix'' as 
\begin{equation}
    C^{-1}(x,y)=  \mtx{0 & i  \\ i & 0}\delta(x-y)\  
\end{equation}
such that $\int dz C(x,z) C^{-1}(z,y)=\delta(x-y)$ is satisfied.
The Dirac bracket is defined by
\begin{multline}
     \{F(x),G(y)\}_D = \{F(x),G(y)\}_P \\-\int dzdw\sum_{i,j=1,2} \{F(x),\phi_i(z)\}_PC^{-1}_{ij}(z,w)\{\phi_j(w),G(y)\}_P\ .
\end{multline}
The Dirac brackets for Majorana fields are computed as
\begin{equation}
\begin{split}
    &\{\Psi(x),\Psi(y)\}_D =\{\Psi^\dagger(x),\Psi^\dagger(y)\}_D =  0\ ,\\
    &\{\Psi(x),\Psi^\dagger(y)\}_D =  - i \delta(x-y)\ . 
\end{split}
\end{equation}
We obtain canonical anti-commutation relations~(\ref{AntiCanRel}) replacing Dirac brackets by anti-commutators as 
$\{\ ,\ \}_D\to -i\{\ ,\ \}$. 

\subsection{Comment on the choice of zweibeins}

In the main text of this paper, we chose zweibeins as in Eq.~(\ref{zwei0}). 
However, this is not a unique choice of local Lorentz frame. We can also consider more general zweibeins $e'{}_\mu^i$ by the local Lorenz transformation of Eq.~(\ref{zwei0}) as
\begin{equation}
\begin{split}
    &\mtx{e'{}^0_\mu \\ e'{}^1_\mu}
    =\mtx{\cosh \theta(t,x) & -\sinh \theta(t,x) \\
    -\sinh \theta(t,x) & \cosh \theta(t,x)}\mtx{e^0_\mu \\ e^1_\mu}\ ,\\
    &\mtx{e'{}_0^\mu \\ e'{}_1^\mu}
    =\mtx{\cosh \theta(t,x) & \sinh \theta(t,x) \\
    \sinh \theta(t,x) & \cosh \theta(t,x)}\mtx{e_0^\mu \\ e_1^\mu}\ ,\\
\end{split}
\end{equation}
where $\theta(t,x)$ is an arbitrary real function. 

Since the Majorana fermion is a representation of the Lorentz group, it is also transformed by the local Lorentz transformation 
\begin{equation}
    \psi=\exp\left(-\frac{1}{2}\theta \gamma^0 \gamma^1\right)\psi'=
    \begin{pmatrix}
\cosh (\theta/2)  & \sinh (\theta/2)\\
\sinh (\theta/2) & \cosh (\theta/2)
\end{pmatrix}\psi'\ .
\end{equation}
The complex variables $\chi$ and $\chi^\dagger$ are transformed as
\begin{equation}
\begin{split}
\chi&=\cosh(\theta/2)\, \chi' -i\sinh (\theta/2)\, \chi'{}^\dagger\ ,\\
\chi^\dagger&=i\sinh (\theta/2)\, \chi'+\cosh(\theta/2)\, \chi'{}^\dagger\ .
\label{LLchi}
\end{split}
\end{equation}

From Eq.~(\ref{AntiCanRel}), anti-commutation relations for variables $\chi$ and $\chi^\dagger$ are given by
\begin{equation}
\begin{split}
\{\chi(x),\chi^\dagger(y)\}&=\frac{1}{\gamma(x)}\delta(x-y)\ ,\\
\{\chi(x),\chi(y)\}&=\{\chi^\dagger(x),\chi^\dagger(y)\}=0\ .
\end{split}
\end{equation}
Therefore, for variables after the local Lorentz transformation, we have anti-commutation relations as
\begin{equation}
\begin{split}
    &\{\chi'(x),\chi'^\dagger(y)\}=\frac{\cosh\theta(x)}{\gamma(x)}\delta(x-y)\ ,\\
    &\{\chi'(x),\chi'(y)\}=-\{\chi'{}^\dagger(x),\chi'{}^\dagger(y)\}=\frac{i\sinh\theta(x)}{\gamma(x)}\delta(x-y)\ .
\end{split}
\end{equation}
The above expressions depend on the choice of zweibeins. 
The anti-commutation relations become canonical only when zweibeins are chosen as in Eq.~(\ref{zwei0}).

When we consider the continuum limit of a spin system, 
the Majorana field defined in Eq.~(\ref{Psi_cnt}) satisfies the canonical anti-commutation relations~(\ref{AntiCanRel}). 
Therefore, when considering the mapping between QFT and spin systems, it is convenient to choose the zweibeins as specified in Eq.~(\ref{zwei0}).

\section{Detailed calculations for exactly solvable model}
\label{detail}

\subsection{Continuum theory}

We will summarize the detailed calculation of the number of the produced particles for the exactly solvable cosmological model~(\ref{tanh}). In this subsection, We consider the continuum theory~(\ref{MajRW}). In the next subsection, we will see that we can proceed almost the same argument even for the discrete theory.

For the following calculation, let us consider the simplest case $a(\eta)=1$. 
Then, the matrix $M_k(\eta)$ defined in Eq.~(\ref{Mcont}) is time-independent and can be diagonalized. As the result, we obtain the analytical expression for mode functions as 
\begin{equation}
 \vec{\phi}_k(\eta) = 
\mtx{u_k\\v_k} e^{-i \epsilon_k\eta}\ ,\quad 
\sigma^x \vec{\phi}_{-k}^\ast(\eta) = 
\mtx{v_k\\u_k} e^{i \epsilon_k\eta}\ ,\quad 
\end{equation}
where we define
\begin{equation}
\epsilon_k=\sqrt{m^2+k^2}\ ,\quad 
\begin{pmatrix}
u_k \\
v_k
\end{pmatrix}
=\frac{1}{\sqrt{2\epsilon_k(\epsilon_k-m)}}
\begin{pmatrix}
-m+\epsilon_k \\
ik
\end{pmatrix}
\label{epsuvdef}
\end{equation}
They satisfy $\epsilon_{-k}=\epsilon_k$, $u_{-k}=u_k^\ast=u_k$, $v_{-k}=v_k^\ast = -v_k$ and  $|u_k|^2+|v_k|^2=u_k^2-v_k^2=1$. One can check orthonormal relations, Eq.~(\ref{orth}). The solution of the Heisenberg equation is written as
\begin{equation}
\begin{split}
 \Psi_k(\eta)&=\gamma_k u_k e^{-i \epsilon_k\eta} + 
\gamma_{-k}^\dagger v_k e^{i \epsilon_k\eta}\ ,\\ 
\Psi_{-k}^\dagger(\eta)&=\gamma_k v_k e^{-i \epsilon_k\eta} + 
\gamma_{-k}^\dagger u_k e^{i \epsilon_k\eta}\ .
\end{split}
\end{equation}
Substituting this expression into the Hamiltonian~(\ref{MajRW}), we have
\begin{equation}
 H=\int^{\pi\ell}_{-\pi\ell} dx \mathcal{H}=\sum_{k\in K} \epsilon_k \gamma_k^\dagger \gamma_k + \textrm{const}\ .
\end{equation}
Therefore, the ground state $|\Omega\rangle$ is defined by $\gamma_k |\Omega\rangle=0$ ($\forall k\in K$).

For the exactly solvable cosmological model~(\ref{tanh}), we can also obtain analytical solutions of mode functions as
\begin{equation}
\begin{split}
    &\vec{\phi}_k = 
    \mtx{ u_k^{(1)}\,
z_-^{i\epsilon_k^{(2)}\Delta\eta/2}F(\alpha_-,\beta_- ,\gamma_-,z_+)
\\ 
v_k^{(1)}\, z_-^{-i\epsilon_k^{(2)}\Delta\eta/2} F(\alpha_+^\ast,\beta_+^\ast ,\gamma_+^\ast,z_+)
}    
z_+^{-i\epsilon_k^{(1)}\Delta\eta/2} \ ,\\
&\sigma^x \vec{\phi}_{-k}^\ast = 
    \mtx{  
v_k^{(1)}\, z_-^{i\epsilon_k^{(2)}\Delta\eta/2} F(\alpha_+,\beta_+,\gamma_+,z_+)
\\
u_k^{(1)}\,z_-^{-i\epsilon_k^{(2)}\Delta\eta/2} F(\alpha_-^\ast,\beta_-^\ast,\gamma_-^\ast,z_+)
}    
z_+^{i\epsilon_k^{(1)}\Delta\eta/2},
\end{split}
\label{modeanaly}
\end{equation}
where $F$ is the Gauss hypergeometric function and we define 
\begin{equation}
\begin{split}
 \epsilon_k^{(i)}&=\epsilon_k|_{m\to ma_i}\ ,\quad
 u_k^{(i)}=u_k|_{m\to ma_i}\ ,\\ v_k^{(i)}&=v_k|_{m\to ma_i}\ ,\quad (i=1,2)\ ,
 \label{u12}
\end{split}
\end{equation}
 i.e., the mass is replaced with the effective mass $ma_i$ in definitions of $\epsilon_k$, $u_k$ and $v_k$ in Eq.~(\ref{epsuvdef}).
Also, we define
\begin{equation}
\begin{split}
&z_\pm = \frac{1}{2}\pm \frac{\tanh \eta/\Delta\eta}{2}\ ,\\
&\alpha_\pm = \frac{i}{2}(\epsilon_k^{(2)}\pm\epsilon_k^{(1)}-ma_1+ma_2)\Delta\eta\ ,\\
&\beta_\pm = 1+\frac{i}{2}(\epsilon_k^{(2)}\pm\epsilon_k^{(1)}+ma_1-ma_2)\Delta\eta\ ,\\
&\gamma_\pm=1\pm i\epsilon_k^{(1)}\Delta\eta\ .
\end{split}
\end{equation}
At the sufficiently early time $\eta\to -\infty$, the mode functions become
\begin{equation}
    \vec{\phi}_k \simeq 
    \mtx{u_k^{(1)}
\\ 
v_k^{(1)}
}    
e^{-i\epsilon_k^{(1)}  \eta}
\ ,\quad
\sigma^x \vec{\phi}_{-k}^\ast \simeq 
    \mtx{v_k^{(1)}
\\ 
u_k^{(1)}
}    
e^{i\epsilon_k^{(1)} \eta}
\ .
\end{equation}
Therefore, 
the ground state at the sufficiently early time is defined by $\gamma_k |\Omega\rangle = 0$ ($\forall k \in K$).
We take $|\Omega\rangle$ as the quantum state and evaluate the expectation value of the number operator.

At the sufficiently late time $\eta\to \infty$, mode functions~(\ref{modeanaly}) become
\begin{widetext}
\begin{equation}
\begin{split}
&\vec{\phi}_k \simeq
    \mtx{ u_k^{(1)}(\frac{\Gamma(\gamma_-)\Gamma(\gamma_--\alpha_--\beta_-)}{\Gamma(\gamma_--\alpha_-)\Gamma(\gamma_--\beta_-)}e^{-i\epsilon_k^{(2)}\eta}
+\frac{\Gamma(\gamma_-)\Gamma(\alpha_-+\beta_--\gamma_-)}{\Gamma(\alpha_-)\Gamma(\beta_-)}e^{i\epsilon_k^{(2)}\eta})
\\ 
v_k^{(1)} (\frac{\Gamma(\gamma_+^\ast)\Gamma(\gamma_+^\ast-\alpha_+^\ast-\beta_+^\ast)}{\Gamma(\gamma_+^\ast-\alpha_+^\ast)\Gamma(\gamma_+^\ast-\beta_+^\ast)}e^{i\epsilon_k^{(2)}\eta}
+\frac{\Gamma(\gamma_+^\ast)\Gamma(\alpha_+^\ast+\beta_+^\ast-\gamma_+^\ast)}{\Gamma(\alpha_+^\ast)\Gamma(\beta_+^\ast)}e^{-i\epsilon_k^{(2)}\eta})
}\ ,\\
    &\sigma^x \vec{\phi}_{-k}^\ast \simeq 
    \mtx{ 
v_k^{(1)} (\frac{\Gamma(\gamma_+)\Gamma(\gamma_+-\alpha_+-\beta_+)}{\Gamma(\gamma_+-\alpha_+)\Gamma(\gamma_+-\beta_+)}e^{-i\epsilon_k^{(2)}\eta}
+\frac{\Gamma(\gamma_+)\Gamma(\alpha_++\beta_+-\gamma_+)}{\Gamma(\alpha_+)\Gamma(\beta_+)}e^{i\epsilon_k^{(2)}\eta})
\\
u_k^{(1)}(\frac{\Gamma(\gamma_-^\ast)\Gamma(\gamma_-^\ast-\alpha_-^\ast-\beta_-^\ast)}{\Gamma(\gamma_-^\ast-\alpha_-^\ast)\Gamma(\gamma_-^\ast-\beta_-^\ast)}e^{i\epsilon_k^{(2)}\eta}
+\frac{\Gamma(\gamma_-^\ast)\Gamma(\alpha_-^\ast+\beta_-^\ast-\gamma_-^\ast)}{\Gamma(\alpha_-^\ast)\Gamma(\beta_-^\ast)}e^{-i\epsilon_k^{(2)}\eta})
}\ ,
\end{split}
\label{phiinf}
\end{equation}
\end{widetext}
where $\Gamma (x)$ denotes the Gamma function. 
On the other hand, natural mode functions at $\eta\to \infty$ are given by
\begin{equation}
    \vec{\Phi}_k \simeq 
    \mtx{u_k^{(2)}
\\ 
v_k^{(2)}
}    e^{-i\epsilon_k^{(2)} \eta} \ ,\quad
\sigma^x \vec{\Phi}_{-k}^\ast \simeq 
    \mtx{v_k^{(2)}
\\ 
u_k^{(2)}
}    e^{i\epsilon_k^{(2)} \eta} \ ,
\label{Phiinf}
\end{equation}
Therefore, using Eqs.~(\ref{Num1}), (\ref{phiinf}) and (\ref{Phiinf}), we have the number of particles produced by the cosmic expansion as
\begin{equation}
\begin{split}
n_k=&\langle \Gamma_k^\dagger \Gamma_k\rangle \\
=&\bigg|u_k^{(2)} v_k^{(1)}\frac{\Gamma(\gamma_+)\Gamma(\gamma_+-\alpha_+-\beta_+)}{\Gamma(\gamma_+-\alpha_+)\Gamma(\gamma_+-\beta_+)}\\
&-v_k^{(2)} u_k^{(1)}\frac{\Gamma(\gamma_-^\ast)\Gamma(\alpha_-^\ast+\beta_-^\ast-\gamma_-^\ast)}{\Gamma(\alpha_-^\ast)\Gamma(\beta_-^\ast)}\bigg|^2
\end{split}
\label{nkanalytic}
\end{equation}
The typical profile of $n_k$ is shown by the black curve in Fig.~\ref{spectrum}.

\subsection{Discrete theory}
\label{exact_disc}

Here, we consider the discrete theory~(\ref{HRWdisc}). 
In case of the flat spacetime, i.e., $a(\eta)=1$, by the similar way as the continuum theory, we can diagonalize the matrix $M_\kappa$ defined in Eq.~(\ref{Mdisc}) and we  have the exact solution for mode functions as 
\begin{equation}
    \vec{\phi}_\kappa(\eta) = \mtx{
u_\kappa\\
v_\kappa
    }e^{-i\epsilon_\kappa \eta}\ ,
\end{equation}
where 
\begin{equation}
    \epsilon_\kappa=\sqrt{z_\kappa^2+y_\kappa^2}\ ,~
    \begin{pmatrix}
u_\kappa \\
v_\kappa
\end{pmatrix}
=\frac{1}{\sqrt{2\epsilon_\kappa(\epsilon_\kappa+z_\kappa)}}
\begin{pmatrix}
\epsilon_\kappa+z_\kappa \\
iy_\kappa
\end{pmatrix}
\label{epsuvdef2}
\end{equation}
and 
\begin{equation}
z_\kappa=\frac{1}{\varepsilon}(1-\cos\kappa-\varepsilon m)\ ,\quad y_\kappa=\frac{1}{\varepsilon}\sin \kappa\ .
\end{equation}
Again the ground state at the early time is defined by $\gamma_\kappa |\Omega\rangle = 0$ ($\forall \kappa \in \mathcal{K}$).
As in Eq.~(\ref{u12}), we again define $u_\kappa^{(i)}$, $v_\kappa^{(i)}$ and $\epsilon_\kappa^{(i)}$ by replacing the mass by the effective mass as $m\to ma_i$.

For the scale factor defined in Eq.~(\ref{tanh}), we again obtain the analytical solution even for the discrete theory.
Note that we can schematically obtain the matrix $M_\kappa(\eta)$ defined in Eq.~(\ref{Mdisc}) 
from that of the continuum theory~(\ref{Mcont}) by changing parameters as
\begin{equation}
    a_i\to a_i-\frac{1-\cos\kappa}{m\varepsilon} \quad (i=1,2)\ ,\quad k\to \frac{\sin \kappa}{\varepsilon}\ .
    \label{paratrans}
\end{equation}
Therefore, the mode function in the discrete theory is given by Eq.~(\ref{modeanaly}) after replacing parameters as in Eq.~(\ref{paratrans}).
Also, the number of produced particles with wavenumber $\kappa$ is given by Eq.~(\ref{nkanalytic}) after changing parameters as in Eq.~(\ref{paratrans}).

\section{On the experimental measurements of the number of produced particles in the expanding Universe}
\label{on_exp_measure}

For the experimental point of view, correlation functions of spin operators would be nice observables.
Here, we show the explicit relation between the number of produced particles $n_k$ and correlation  functions of spin operators.

At the sufficiently late time, the Hamiltonian is static and we can diagonalize it by the Bogoliubov transformation:
\begin{equation}
\begin{split}
    &\Gamma_\kappa=e^{i\epsilon_\kappa^{(2)}\eta}(u_\kappa^{(2)} c_\kappa(\eta)  - v_\kappa^{(2)} c_{-\kappa}^\dagger(\eta) )\ ,\\
    &\Gamma_\kappa^\dagger=e^{-i\epsilon_\kappa^{(2)}\eta}(u_\kappa^{(2)} c_\kappa^\dagger(\eta)  + v_\kappa^{(2)} c_{-\kappa}(\eta) )\ .
\end{split}
\end{equation}
See appendix~\ref{exact_disc} for the definition of $u_\kappa^{(i)}$, $v_\kappa^{(i)}$ and $\epsilon_\kappa^{(i)}$ ($i=1,2$). 
The expectation value of the number operator is written as
\begin{equation}
\begin{split}
    n_k=&\langle \Gamma_k^\dagger \Gamma_k\rangle\\ = & (u_\kappa^{(2)})^2 \langle c_\kappa^\dagger c_\kappa \rangle - (v_\kappa^{(2)})^2 \langle c_{-\kappa} c_{-\kappa}^\dagger \rangle\\
    &+ u_\kappa^{(2)} v_\kappa^{(2)} ( \langle c_{-\kappa} c_{\kappa} \rangle -\langle c_\kappa^\dagger c_{-\kappa}^\dagger \rangle) \ ,
\end{split}
\end{equation}
where $\langle \cdots \rangle$ is the expectation value with respect to the initial ground state $|\Omega\rangle$. Using fermionic operators in the position space $c_j$ and $c_j^\dagger$, we have
\begin{equation}
\begin{split}
    n_k=\frac{1}{L}\sum_{j,k=1}^L e^{i\kappa (j-l)} &\bigg[(u_\kappa^{(2)})^2 \langle c_j^\dagger c_l \rangle - (v_\kappa^{(2)})^2 \langle c_{j} c_{l}^\dagger \rangle 
    \\&+ u_\kappa^{(2)} v_\kappa^{(2)} ( \langle c_{j} c_{l} \rangle -\langle c_j^\dagger c_{l}^\dagger \rangle)\bigg]     \ .
    \label{nkcorr}
\end{split}
\end{equation}
Therefore, $n_k$ is computed from two point functions of fermionic operators. 
In terms of spin operators $\sigma^\pm_j$ and $\sigma^z_j$, the two point function is written as
\begin{equation}
\begin{split}
    &\langle c_j^\dagger c_l \rangle = \langle \sigma^-_{j} \sigma^+_l \prod_{r\in (j,l)} \sigma^z_r \rangle,\\
    &\langle c_j c_l^\dagger \rangle = (-1+2\delta_{jl})\langle \sigma^+_{j} \sigma^-_l \prod_{r\in (j,l)} \sigma^z_r\rangle\ ,\\
    &\langle c_j c_l \rangle = \textrm{sgn}(j-l) \langle \sigma^+_{j}  \sigma^+_l \prod_{r\in (j,l)} \sigma^z_r\rangle\ ,\\
    &\langle c_j^\dagger c_l^\dagger \rangle = -\textrm{sgn}(j-l) \langle \sigma^-_{j}  \sigma^-_l \prod_{r\in (j,l)} \sigma^z_r\rangle\ ,
\end{split}
\label{corrfuncs}
\end{equation}
where $\textrm{sgn}(n) = 1,0,-1$ for $n>0$, $n=0$ and $n<0$, respectively. 
We define
\begin{equation}
    (j,l)=
    \begin{cases}
    \{j+1,j+2,\cdots,l-1\} & (j\leq l)\\
    \{l+1,l+2,\cdots,j-1\} & (j> l)
    \end{cases}\ ,
\end{equation}
i.e., the set of sites between $j$- and $l$-th sites (which does not include its endpoints). When $|j-l|\leq 1$, $(j,l)$ is the empty set. When $(j,l)$ is empty, we define $\prod_{r\in (j,l)} \sigma^z_r=1$.
The right-hand sides of Eq.~(\ref{corrfuncs}) are $(|j-l|+1)$-point functions. The product of $\sigma^z$'s in multi-point functions is rewritten as
\begin{equation}
    \prod_{r\in (j,l)} \sigma^z_r = \prod_{r\in (j,l)} \sigma^z_r \, e^{i\pi N} e^{i\pi N} 
    =\prod_{r\notin (j,l)} \sigma^z_r \, e^{i\pi N} \ ,
\end{equation}
where $e^{i\pi N}$, defined in Eq.~(\ref{oddity}), measures the oddity of the total number of fermions. For the initial ground state, we  have $e^{i\pi N}=1$. Thus, we can rewrite Eq.~(\ref{corrfuncs}) as
\begin{equation}
\begin{split}
    &\langle c_j^\dagger c_l \rangle = -\langle \sigma^-_{j} \sigma^+_l \prod_{r\notin (j,l), r\neq j,l} \sigma^z_r \rangle\ ,\\
    &\langle c_j c_l^\dagger \rangle = \langle \sigma^+_{j} \sigma^-_l \prod_{r\notin (j,l), r\neq j,l} \sigma^z_r\rangle\ ,\\
    &\langle c_j c_l \rangle = \textrm{sgn}(j-l) \langle \sigma^+_{j}  \sigma^+_l \prod_{r\notin (j,l), r\neq j,l} \sigma^z_r\rangle\ ,\\
    &\langle c_j^\dagger c_l^\dagger \rangle = -\textrm{sgn}(j-l) \langle \sigma^-_{j}  \sigma^-_l \prod_{r\notin (j,l), r\neq j,l} \sigma^z_r\rangle\ .
\end{split}
\label{corrfuncs2}
\end{equation}
They are $(L-|j-l|+1)$-point functions. 
By using the above expressions, we can reduce the number of operators in multi-point functions when $|j-l|> L/2$. 
Therefore, once we can measure $2$-, $3$-, $\ldots$, $L/2$-point functions in Eqs.~(\ref{corrfuncs}) or (\ref{corrfuncs2}), we can compute the number of the produced particle from Eq.~(\ref{nkcorr}).

\begin{figure}[tbh]
  \centering
\subfigure[$L=64$]
 {\includegraphics[scale=0.35]{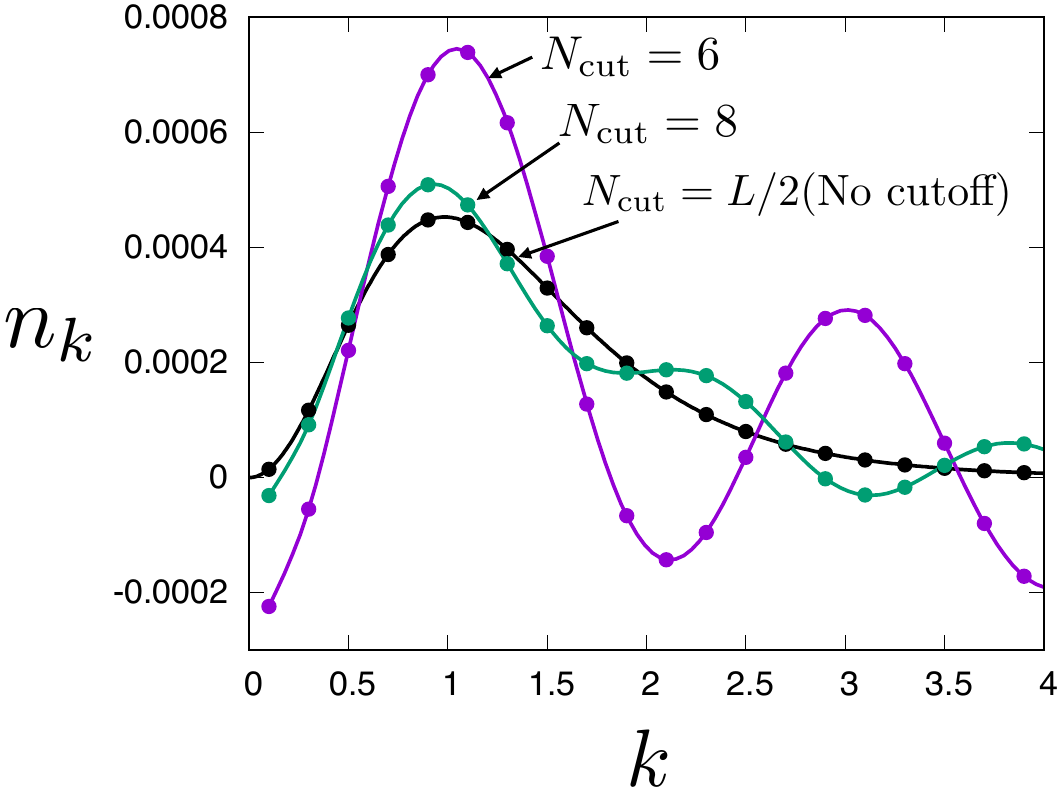}
  }
  \subfigure[$L=128$]
 {\includegraphics[scale=0.35]{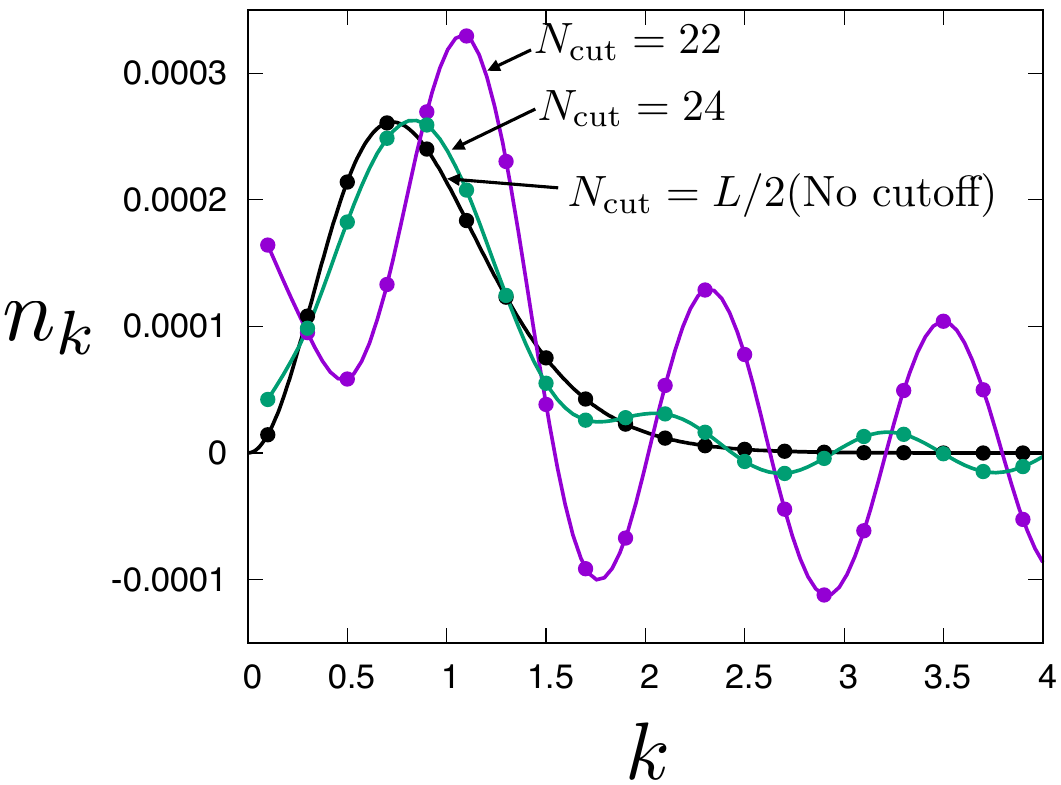}
  }
 \caption{$N_\textrm{cut}$-dependence of Eq.~(\ref{Ncut}) for $L=64$ and $L=128$. 
 Parameters are $m=\Delta\eta=a_1=1$, $a_2=2, \ell=5$ and $\eta=5$.}
 \label{Ncut_dep}
\end{figure}

The measurement of multi-point functions with many inserted operators would be difficult. 
We can expect that correlations becomes small for sufficiently separated points on the ring. 
Eq.~(\ref{nkcorr}) would be approximated as 
\begin{equation}
    \begin{split}
    n_k\simeq &\frac{1}{L}\sum_{\substack{|j-l|\leq N_\textrm{cut} \textrm{ or}\\ L-|j-l|\leq N_\textrm{cut}}}
    e^{i\kappa (j-l)} \bigg[(u_\kappa^{(2)})^2 \langle c_j^\dagger c_l \rangle\\& - (v_\kappa^{(2)})^2 \langle c_{j} c_{l}^\dagger \rangle + u_\kappa^{(2)} v_\kappa^{(2)} ( \langle c_{j} c_{l} \rangle -\langle c_j^\dagger c_{l}^\dagger \rangle)\bigg]     \ ,
    \label{Ncut}
    \end{split}
\end{equation}
i.e., correlation functions of two points further than the cutoff $N_\textrm{cut}$ are set to zero by hand. Figure~\ref{Ncut_dep} shows $N_\textrm{cut}$-dependence of above expression for $L=64$ and $L=128$. 
We use the same parameter as in Fig.~\ref{spectrum}. We also set the time coordinate as $\eta=5$ since 
Eq.~(\ref{Ncut}) is time-dependent when there is a cutoff. 
For $N_\textrm{cut}\gtrsim 8$ ($L=64$) and $N_\textrm{cut}\gtrsim 24$ ($L=128$), 
we see qualitative agreement with results with no cutoff. 

\bibliography{refs}

\end{document}